\documentclass[twocolumn]{aastex631} 

\usepackage{amsmath}
\usepackage{CJK}
\usepackage[]{natbib}
\usepackage[normalem]{ulem}
\usepackage{booktabs}
\hypersetup{
   colorlinks,
   linkcolor={blue!88!black!80},
   citecolor={blue!88!black!80},
   urlcolor={blue!88!black!80}}

\newcommand{\kms}{km s\ensuremath{^{-1}}}

\newcommand{\RNum}[1]{\uppercase\expandafter{\romannumeral #1\relax}}

\newcommand {\hb}{\ifmmode \mathrm{ H}\beta \else H$\beta$\fi}
\newcommand {\ha}{\ifmmode \mathrm{ H}\alpha \else H$\alpha$\fi}
\newcommand {\hr}{\ifmmode \mathrm{ H}\gamma \else H$\gamma$\fi}

\begin{document}
\begin{CJK}{UTF8}{gbsn}

\title{AGN Continuum Reverberation Mapping Based on Zwicky Transient Facility Light Curves}

\author[0000-0001-8416-7059]{Hengxiao Guo (郭恒潇)}
\affiliation{Key Laboratory for Research in Galaxies and Cosmology, Shanghai Astronomical Observatory, Chinese Academy of Sciences, 80 Nandan Road, Shanghai 200030, People's Republic of China}
\affiliation{Department of Physics and Astronomy, 4129 Frederick Reines Hall, University of California, Irvine, CA, 92697-4575, USA}

\author[0000-0002-3026-0562]{Aaron J. Barth}
\affiliation{Department of Physics and Astronomy, 4129 Frederick Reines Hall, University of California, Irvine, CA, 92697-4575, USA}

\author[0000-0002-2052-6400]{Shu Wang}
\affiliation{Department of Physics \& Astronomy, Seoul National University, Seoul 08826, Republic of Korea}

\begin{abstract}
We perform a systematic survey of active galactic nucleus (AGN) continuum lags using $\sim$3 day cadence $gri$-band light curves from the Zwicky Transient Facility. We select a sample of 94 type~1 AGN at $z<0.8$ with significant and consistent inter-band lags based on the interpolated cross-correlation function method and the Bayesian method {\tt JAVELIN}. Within the framework of the ``lamp-post'' reprocessing model, our findings are:  1) The continuum emission (CE) sizes inferred from the data are larger than the disk sizes predicted by the standard thin disk model;  2) For a subset of the sample, the CE size exceeds the theoretical limit of the self-gravity radius (12 lt-days) for geometrically thin disks; 3) The CE size scales with continuum luminosity as $R_{\mathrm{CE}} \propto L^{0.48\pm0.04}$ with a scatter of 0.2 dex, analogous to the well-known radius-luminosity relation of broad \hb. These findings suggest a significant contribution of diffuse continuum emission from the broad-line region (BLR) to AGN continuum lags. We find that the $R_{\mathrm{CE}}-L$ relation can be explained by a photoionization model that assumes $\sim$23\% of the total flux comes from the diffuse BLR emission. In addition, the ratio of the CE size and model-predicted disk size anti-correlates with the continuum luminosity, indicative of a potential non-disk BLR lag contribution evolving with luminosity. Finally, a robust positive correlation between CE size and black hole mass is detected.

\end{abstract}

\section{Introduction} \label{sec:intro} 
Accretion disks around the supermassive black holes (SMBHs) in the centers of active galactic nuclei (AGN) are naturally formed by infalling gas that sinks into the galaxy's central gravitational potential. A geometrically thin, optically thick accretion disk \citep[i.e., Shakura-Sunyaev disk or SSD,][]{Shakura73} that emits continuum radiation across UV-optical wavelengths is thought to be the standard mode of accretion in powerful AGN. However, real accretion disks are more complicated and heterogeneous: the disk structure is accretion rate ($\dot{m}={L}/{L}_\mathrm{Edd}$) dependent and the cold, radiatively efficient SSD mode applies to moderately sub-Eddington accretion ($\dot{m}<1$). Disks with accretion rates higher than the Eddington limit or much below it can yield different disk structures, such as the slim disk ($\dot{m} \gtrsim$ 1) \citep{Abramowicz88,Abramowicz13} and the advection-dominated accretion flow ($\dot{m} \ll$ 1) \citep{Narayan95,Yuan14}. Additionally, strong magnetic fields are likely to affect the disk physics and structure, for example, by changing the spectral energy distribution and radial temperature profile \citep[e.g.,][]{Slone12,Laor14,Li19,Sun19}. 

According to the widely used SSD model, the UV-to-optical emission from the disk is a superposition of multi-temperature blackbody radiation. The standard accretion disk exhibits a temperature gradient with its maximum effective temperature ($T_\mathrm{ eff} \sim 10^{5-6}$ K) near the center and a decrease with  disk radius ($R$) following $T_\mathrm{ eff} \propto R^{-3/4}$. Using Wien's law to associate $T_\mathrm{eff}$ with the corresponding blackbody peak wavelength, the radius of peak emission at wavelength $\lambda$ scales as $R_{\lambda} \propto \lambda^{\beta}$, where $\beta=4/3$ for the SSD model.

Reverberation mapping (RM) is an effective tool to indirectly resolve the disk structure through light echoes \citep{Blandford82,Peterson93,Cackett21}. The variability in AGN is usually assumed to follow the lamp-post X-ray reprocessing model \citep{Krolik91,Cackett07}, in which the disk UV/optical emission is a product of X-ray reprocessing from a corona above the inner disk region. The lower-energy emission is produced in relatively more extended regions of the accretion disk leading to an observed time lag, for example between UV and  optical wavelengths, which corresponds to the distance between two emitting annuli. However, this reprocessing scenario has been challenged in many aspects including the deficit of the energy budget, the unclear correlation between X-ray and UV/optical variations, timescale-dependent color variations, and the disk size problem \citep[e.g.,][and references therein]{Cai18}, and other models have been proposed for the origin of UV/optical variations such as the inhomogeneous disk model in which variability is driven by temperature fluctuations in the disk \citep{Dexter11,Cai18,Sun20a,Neustadt22}.

Assuming that the observed UV/optical continuum variations from the accretion disk are driven by X-ray reprocessing from the corona, the disk sizes at different radii can be easily inferred by measuring the corresponding continuum lags with continuum RM techniques \citep[for a review see][]{Cackett21}. Then, the inter-band lags are predicted to be 
\begin{equation}\label{eq:tau}
    \tau = \frac{R_{\lambda_{0}}}{c}\Bigg[\Bigg(\frac{\lambda}{\lambda_{0}}\Bigg)^{\beta}-1 \Bigg],
\end{equation}
where $\tau$ is the time lag between emission from two different annuli emitting at wavelengths $\lambda$ and $\lambda_{0}$. This model allows us to estimate the disk size at any wavelength and compare it to the theoretical predictions given $R_{\lambda_{0}}$ and $\beta$ constrained from the observations.

Tight inter-band correlations of continuum emission were detected in early reverberation mapping studies \citep[e.g.,][]{Clavel91,Krolik91}, and the first robust positive wavelength-lag relation was found in UV observations of NGC 7469 \citep{Wanders97,Collier98,Kriss00}. Later on, mounting evidence was gradually reported to support this relation in various AGN \citep[e.g.,][]{Collier01,Sergeev05}. Recently, thanks to large investments of time on both space-based and ground-based telescopes, intensive monitoring programs have successfully resolved continuum reverberation lags from X-ray to near-infrared (NIR) wavelengths in a growing number of sources  \citep[e.g.,][]{Mchardy14,Shappee14,Edelson15,Fausnaugh16,Edelson17,Fausnaugh18,Edelson19,Cackett18,Cackett20,HernandezSantisteban20,Lobban20,Vincentelli21,Kara21}. As a complement to intensive single-object campaigns, large time-domain surveys are capable of exploring the continuum reverberation properties of large samples of AGN \citep{Jiang17,Mudd18,Yu20,Homayouni19,Jha21}, although typically with data of lower cadence and limited wavelength coverage. 

Key results of these continuum RM campaigns can be summarized as follows: 
\begin{enumerate}
    \item UV-optical-NIR lags generally follow the $\tau \propto \lambda^{4/3}$ dependence predicted by a thin disk reprocessing model, but typically with a normalization of $\sim$3 times larger than the prediction, in agreement with microlensing measurements \citep[e.g.,][]{Morgan10}.
    
    \item Lags around the Balmer jump (3647 \AA), and possibly also the Paschen jump (8206 \AA), are generally systematically longer than expected based on an extrapolation of the other UV/optical lags \citep[e.g.,][]{Fausnaugh16,Cackett18}.
    
    \item Continuum lags are correlated with other AGN properties, such as continuum luminosity and black hole mass. However, their consistency with SSD predictions is still unclear.
\end{enumerate}

Explanations for the longer-than-expected inter-band lags have been proposed, e.g., increasing the predicted continuum lag by modifying the SSD model \citep{Starkey17,Cai18,Sun20a}, altering the reprocessing geometry \citep{Gardner17,Kammoun21}, incorporating the effects of winds \citep{Li19,Sun19}, considering the AGN internal extinction \citep{Gaskell17}, and including other continuum components in addition to disk emission  \citep[e.g., diffuse BLR emission,][]{Lawther18,Hall18,Korista19,Chelouche19,Netzer21,Cackett21b,Guo22}. 

Our goal for this work was to construct a large new sample of type 1 AGN with robust continuum lag detections, using data from the Zwicky Transient Facility (ZTF) survey, to further explore the relationship between continuum emission size and other AGN properties. Our new sample allows us to explore theoretically predicted relations, e.g., the $\tau-L$ relation \citep{Netzer21}, the $\tau-M_\mathrm{ BH}$ relation \citep{Yu20}, and the dependence of the disk-size discrepancy (between observation and SSD prediction) on luminosity \citep{Li21}.
In \S \ref{sec:sample}, we describe the data and preliminary sample selection. In \S \ref{sec:lag}, we use the interpolated cross-correlation function (ICCF) and {\tt JAVELIN} methods to measure continuum lags among the $gri$ bands, estimate the disk sizes, and select subsamples having the best-quality lag measurements for further analysis. Our main results and discussions are given in \S \ref{sec:results}. Finally, we present our conclusions in \S \ref{sec:con}.

\section{Data and Sample Selection} \label{sec:sample}
\subsection{ZTF Survey}

ZTF, an automated time-domain survey, utilizes the 1.2 m (48 inch) Schmidt telescope at Palomar Observatory and a 576-megapixel camera with 47 deg$^2$ field of view \citep{Bellm19}. It has covered the entire sky visible from Palomar (Dec $> -30^{\circ}$, $25,000-30,000$ deg$^2$) since March 2018 (MJD $\ge$ 58194), with a cadence of $3-4$ days in the \emph{gri} bands. The ZTF data has a 5$\sigma$ detection limit of $g=$ 20.8, $r=$~20.6, $i=$ 19.9 mag (AB system) with 30s exposures \citep{Masci19}. Light curves are constructed using measurements from the calibrated single-exposure PSF-fit photometry\footnote{ZTF data can be accessed via {\url{https://irsa.ipac.caltech.edu/docs/program_interface/ztf_lightcurve_api.html}}}. In this work, we adopted the ZTF Data Release 7 (DR7) light curves with a baseline of 4 (2) years in the $gr$ ($i$) bands.

\begin{deluxetable}{lr}[htbp]
\caption{Initial sample selection } \label{tab:selection}
\tablewidth{0pt}
\tablehead{
\colhead{Selection} &
\colhead{Objects}
}
\startdata
Million quasar catalog  &1,562,836 \\
1) Dec $>$ $-$30 deg & 1,374,722  \\
2) Spectroscopically confirmed type 1 AGN/QSO & 814,622 \\
3) $R$ $<$17 mag  & 9,977 \\
4) $N$(epochs) $>$ 20 for $gri$ light curves  & 4,255\\
5) ICCF $r_{\mathrm{max,gr}}$ and $r_{\mathrm{max,gi}} > 0.8$ & 559\\
6) Redshift $z$ $<$ 0.8 & 455 \\
\enddata
\end{deluxetable}

\begin{figure*}
    \centering
    \includegraphics[width=18.cm]{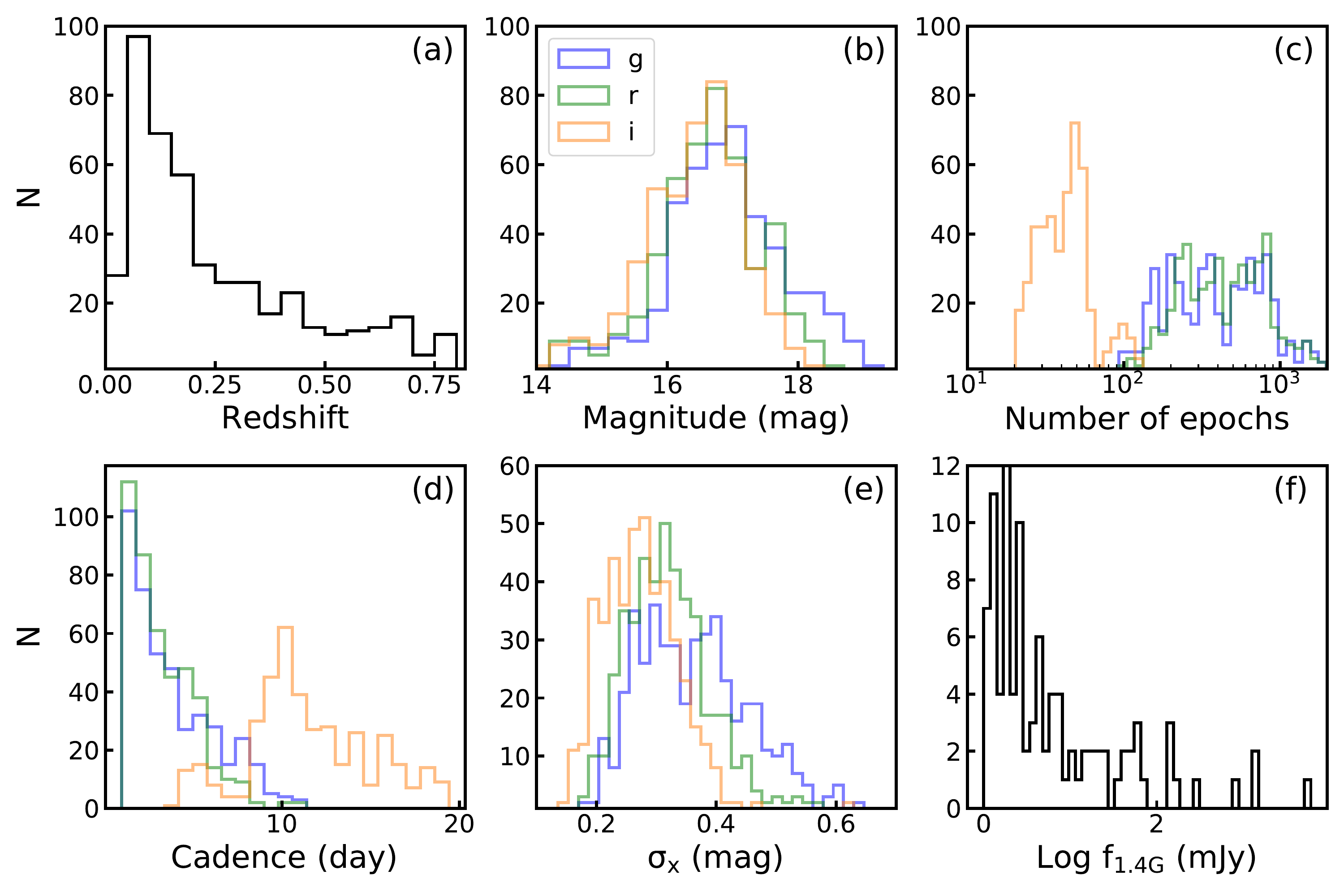}
    \caption{Properties of the initial sample. Panels (a) and (b) show the redshift and mean magnitude distributions, respectively. Different colors represent different bands as shown in the label. Panels (c) to (e) display the distributions of number of epochs, cadence, and square root of excess variance of the light curves. Panel (f) shows the distribution of radio flux density. }
    \label{fig:basic}
\end{figure*}

\subsubsection{Sample Selection}\label{sec:sample1}

In order to identify a large sample of AGN having ZTF data suitable for lag measurement, we started with the Million Quasar Catalog \citep[MQC, version 7.2 released on 30 April 2021,][]{Flesch21}. This catalog totally includes 1,573,824 AGN including 29,666 type 1 QSOs/AGN, 703,348 quasar candidates, and some type~2 objects and blazars. Our selection steps and the sample size resulting from each step are summarized in Table \ref{tab:selection}. We first selected spectroscopically confirmed type 1 QSOs/AGN from the MQC at declination $>-30^\circ$ with $R$-band magnitude brighter than 17 mag to guarantee a relatively high S/N of the light curve for lag detection. These type 1 QSOs/AGN\footnote{In the following, we use AGN generically and do not distinguish the bright QSOs and less-luminous AGN.} are selected by requiring the {\tt Type} parameter to be ``Q'' or ``A''. This step results in 9,977 type 1 AGN selected from the MQC.

We then searched for ZTF DR7 light curves for each object and required the number of good epochs of all of the \emph{g}, \emph{r}, and \emph{i} bands to be at least 20. In this step, 4,255 objects were selected. To ensure that the data in our light curves are unaffected by  clouds, moonlight contamination, or bad seeing, we require {\tt catflags} $= 0$ in our selection of data points. Next, we restricted our sample to those objects that display the most significant interband correlations, which will give the most robust lag measurements. We measured continuum lags using the ICCF method for this sample of 4,255 AGN (see \S \ref{sec:lag} for details of the lag measurement), and calculated the maximum correlation coefficient $r_\mathrm{ max}$ from the ICCF of the inter-band light curves, which is an indicator of the strength of the correlation. We required both $r_{\mathrm{max,gr}}$ and $r_{\mathrm{max,gi}}$ to be larger than 0.8, indicating robust correlations among three-band light curves.

Finally, only low-redshift objects with $z<$ 0.8 are considered in the following work, because: 1) the measurements of black hole (BH) mass can be uniformly derived from the broad \hb\ line; 2) high-redshift objects tend to be relatively faint with low-S/N light curves; 3) at higher $z$, light curve features will be further stretched by time dilation, making lag detection more challenging. This yields 455 AGN as our initial sample.

Figure \ref{fig:basic} illustrates the basic properties of these 455 AGN. Panel (a) shows that the redshift distribution is peaked around 0.05 with $\sim$70\% of objects at $z < $ 0.3. Panel (b) displays the magnitude distributions in the \emph{gri} bands, measured as the mean of each object's magnitude from the full ZTF light curves. Although we have performed an $R$-band brightness cut in sample selection, about 34\% of the sample in ZTF-$r$ (as well as 56\% in ZTF-$g$ and 21\% in ZTF-$i$) are still fainter than 17 mag because of AGN variability, differences between filters, and different photometric methods between MQC data sources. Panels (c) and (d) describe the epochs and average cadence (including seasonal gaps) of ZTF light curves. As the $i$-band photometry is a bonus contributed by the time-limited ZTF collaboration survey, its cadence and temporal coverage (spanning only the first two years) are lower than those of the other bands, yet this shorter duration was still long enough to obtain significant lags for many objects. Panel (e) presents the square root of the light curve excess variance ($\sigma_\mathrm{x}^2$) following \citet{Edelson02}. This evaluates the rms variation amplitude in the light curve over and above the amount expected from the measurement uncertainties on each data point. The median values of $\sigma_\mathrm{x}$ are 0.36, 0.31, 0.27 mag for the $g$, $r$, and $i$ bands respectively, indicating that the variability amplitude decreases with increasing wavelength as expected. Panel (f) exhibits 99 radio-detected ($>$ 1~mJy) objects by cross-matching with FIRST at observed 1.4~GHz \citep{Becker95}. Nine of them have flux density $f_\mathrm{ 1.4G}$ larger than 100 mJy. Other objects either have radio flux density less than 1 mJy or are in fields not covered by the FIRST survey.

\begin{deluxetable}{lr}[htbp]
\caption{BH mass information for the initial sample} \label{tab:mass}
\tablewidth{0pt}
\tablehead{
\colhead{Origins} &
\colhead{Number}
}
\startdata
1) SDSS DR14 Quasar catalog  & 130  \\
2) SDSS DR7 broad-line AGN catalog & 167 \\
3) PG quasar catalog & 38\\
4) LAMOST DR5 & 74 \\ 
5) RM & 20\\
6) SDSS DR16, LAMOST DR6, and literature & 93\\
\hline
Without BH mass & 79\\
\enddata
\end{deluxetable}

\subsubsection{BH Mass Estimates}
In an attempt to explore the correlations between lag and luminosity, and lag and BH mass, we collected the monochromatic continuum luminosities at 5100 \AA\ and single-epoch virial BH masses or reverberation BH masses. All BH masses were measured from the broad \hb\ line. The origins of this information are fairly heterogeneous. For some objects the data were obtained from previous literature that summarized the continuum luminosity and BH mass from spectroscopic surveys or RM campaigns, and for the remainder of the sample we carried out measurements on existing spectra from the Sloan Sky Digital Survey (SDSS), Large Sky Area Multi-Object Fiber Spectroscopic Telescope (LAMOST), or digitized high S/N spectra from previous work. 

We first cross-matched our initial sample of 455 AGN with four catalogs: SDSS DR14 \citep{Rakshit20}, LAMOST DR1 to DR5 \citep{Ai16,Dong18,Yao19}, the catalog of \citet{Liu19} (L19), and the catalog of \citet{Shangguan18} (S18), which include monochromatic luminosities at 5100 \AA\ and BH mass measurements of low-redshift AGN. \citet{Rakshit20} (R20 hereafter) measured the spectral properties for a total of 526,265 quasars in SDSS DR14. They performed global spectral fitting for each spectrum using {\tt PyQSOFit} \citep{Guo18}, and listed the continuum luminosity at 5100 \AA\ and BH masses based on the single-epoch spectrum of \hb\ following the empirical mass calibration from \citet{Vestergaard06} (VP06 hereafter). However, SDSS DR14 defines a quasar as an object with a luminosity $M_{i,z=2} < -20.5$ and either displaying at least one emission line with a full-width at half maximum (FWHM) $>$ 500 \kms\ or having interesting/complex absorption features. This criterion will inevitably reject a large fraction of type~1 AGN at the low-luminosity end and also some potential AGN with very narrow broad components. To address this issue, L19 constructed a more comprehensive and uniform sample of 14,584 broad-lined AGN at $z <$ 0.35 from SDSS DR7. This is a good complement to the DR14 quasar catalog (DR14Q) for the low-redshift AGN, although the BH masses are based on a different calibration \citep{Greene05}. On the other hand, due to the bright magnitude limits, i.e., $m_i$ $>$ 15 mag, very bright nearby AGN can be excluded. To complement the AGN at the bright end, we also included the low-redshift quasar catalog from S18 who listed BH masses of 87 Palomar-Green quasars based on the VP06 calibration. 

LAMOST spectroscopically confirmed 43,109 quasars from 2012 to 2017 \citep[DR1 to DR5,][]{Ai16,Dong18,Yao19}, although half of them have been reported by SDSS. They conducted a local spectral fitting for the \hb\ region following \citet{Dong08}, and estimated the virial black hole mass with the VP06 calibration. 

In addition, we also collected all of the \hb-based BH masses from AGN RM campaigns, which includes $\sim$150 unique objects \citep[e.g.,][]{Peterson04,BentzKatz15,Du15,Barth15,Grier17,U21}.

Since the publication of these works, more AGN have been discovered but without spectral properties reported, e.g., the newest data releases of SDSS (DR17) and LAMOST (DR6). We cross-matched our initial sample with the SDSS and LAMOST databases to download the spectra which are not included in SDSS DR14 and LAMOST DR1 to DR5. We performed local spectral fitting for the \hb\ region (4640$-$5100 \AA) with {\tt PyQSOFit} following \citet{Shen19}, and calculated the BH masses with VP06. 

For 117 AGN without cross-matched BH mass in these catalogs, we searched previous literature and found that the continuum luminosity and BH mass of eight objects were directly reported, and 30 AGN spectra were available \citep[e.g.,][]{Bade95,Engels98,Wei99}. The spectra were digitized from their figures with {\tt WebPlotDigitizer}\footnote{\url{https://automeris.io/WebPlotDigitizer/}} and refitted with {\tt PyQSOFit} to obtain estimates of the BH masses. These 30 spectra, which  include the \hb-[\ion{O}{3}] region, were carefully selected by requiring a line S/N of \hb\ $>$ 10~pixel$^{-1}$. A few other available spectra were rejected due to the low spectral S/N, a very weak \hb\ line, \hb\ approaching the edge of the observed spectra range, telluric absorption, or spectra being displayed as normalized spectral flux. We searched for BH masses from the literature compiled by MQC and we refer readers to the {\tt Ref\_Name} tag in the MQC for further details. 

As listed in Table \ref{tab:mass}, 376 of 455 AGN have continuum luminosity at 5100 \AA\ and \hb-based BH mass collected from previous catalogs, literature, or measured as part of this work. The number of BH masses obtained from each method is also listed in Table \ref{tab:mass}. Some objects are duplicated in different catalogs, and we adopt the average BH masses (also continuum luminosity) as our fiducial values. Although RM masses are expected to be more accurate than single-epoch masses, only 3 and 2 reverberation-mapped objects are included in our parent and core samples in \S \ref{sec:parent_core}, with negligible influence on our results. In addition, we note that R20, L19, and S18 report the continuum luminosity by subtracting the host contribution determined via spectral decomposition, while LAMOST catalogs only list the total continuum luminosity due to the relatively low spectral S/N. R20 performed a comparison of total and AGN-only continuum luminosity at 5100 \AA\ for SDSS DR7Q (see their Figure 6), and suggested that the mean difference is $\sim$0.05 dex (comparable to the uncertainty), decreasing with increasing luminosity. This indicates that the influence on our results from different measures of continuum luminosity in different subsamples is limited.

\begin{figure*}
    \centering
    {{\includegraphics[width=8.5cm]{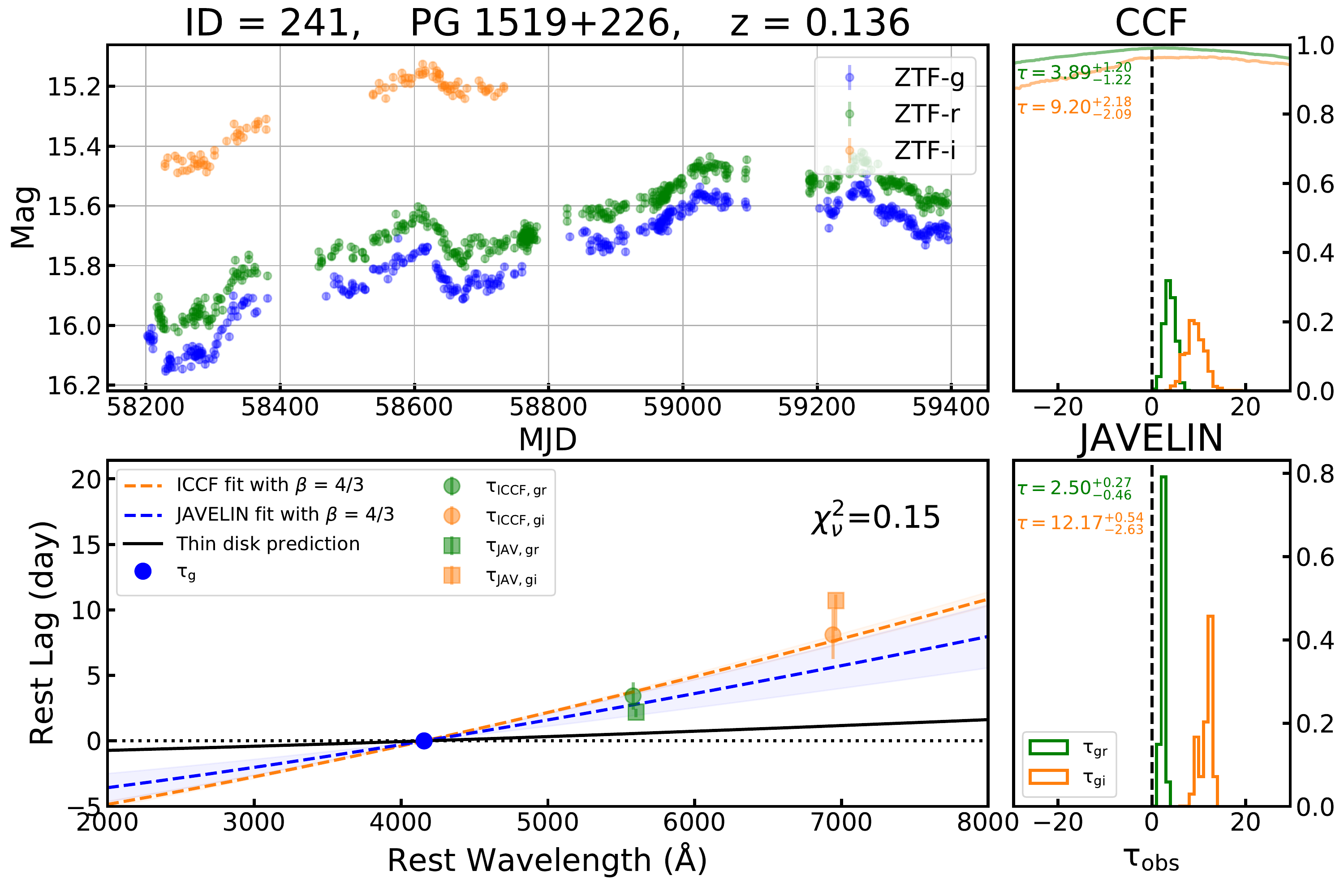} }}%
    {{\includegraphics[width=8.5cm]{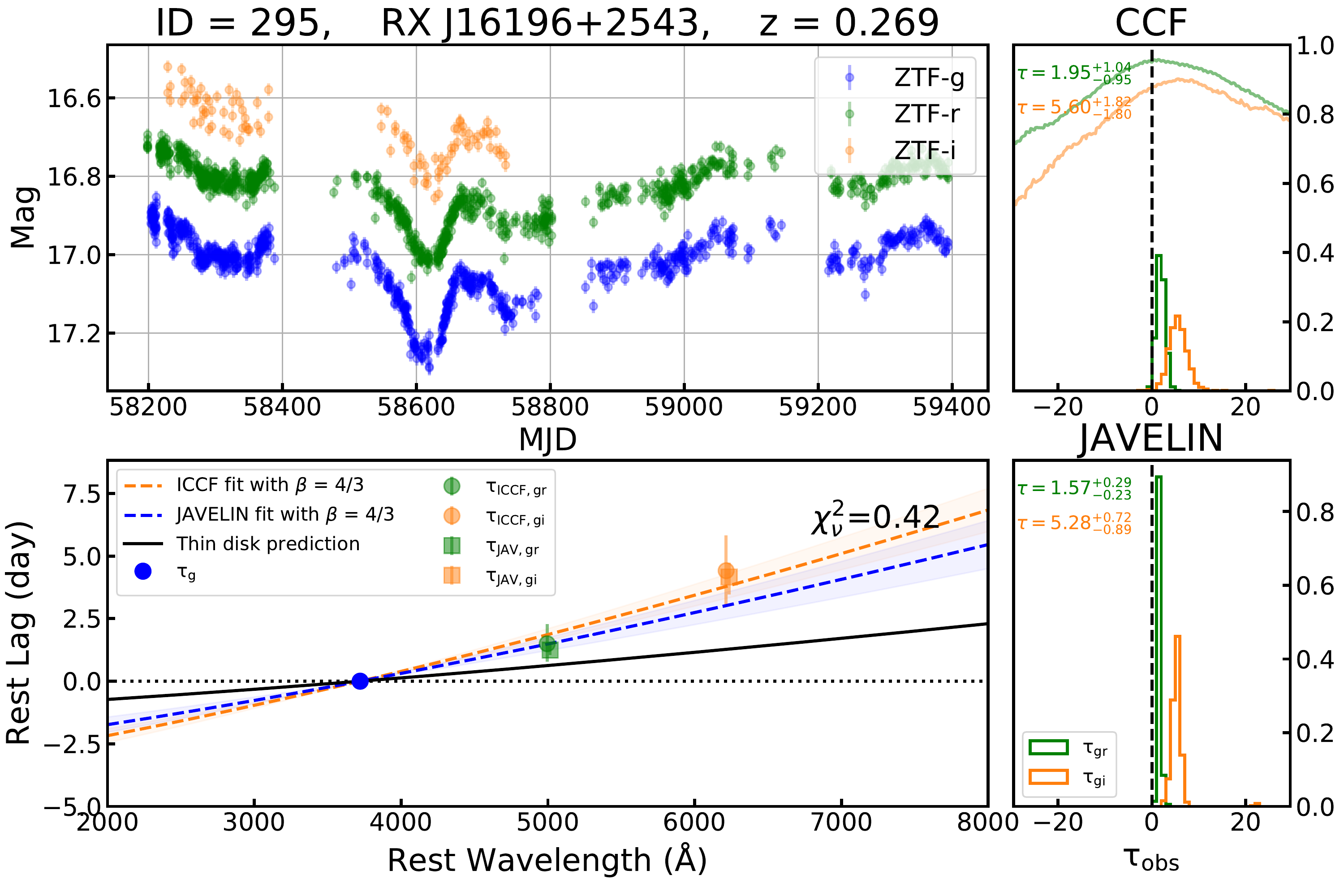} }}%
    {{\includegraphics[width=8.5cm]{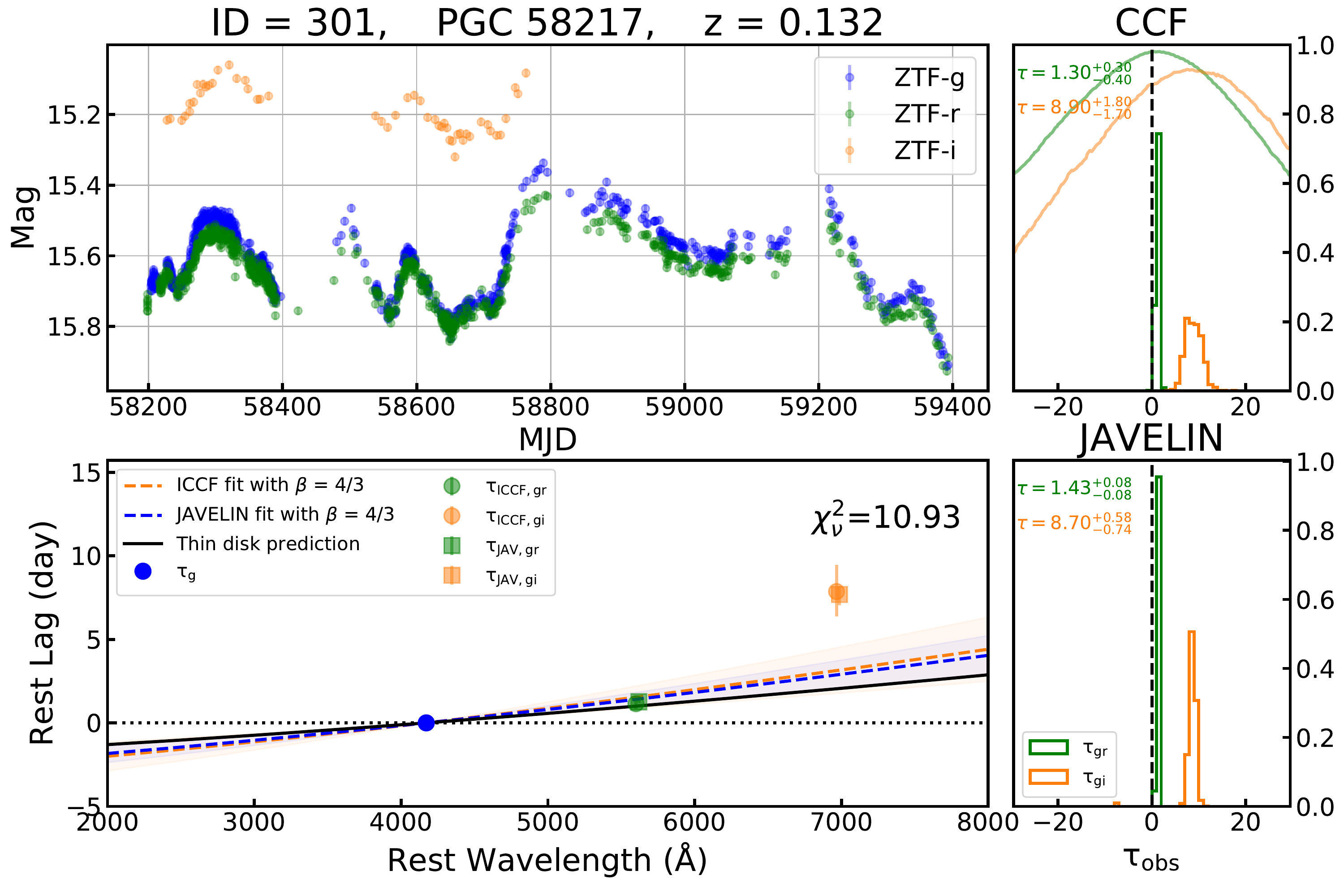} }}%
    {{\includegraphics[width=8.5cm]{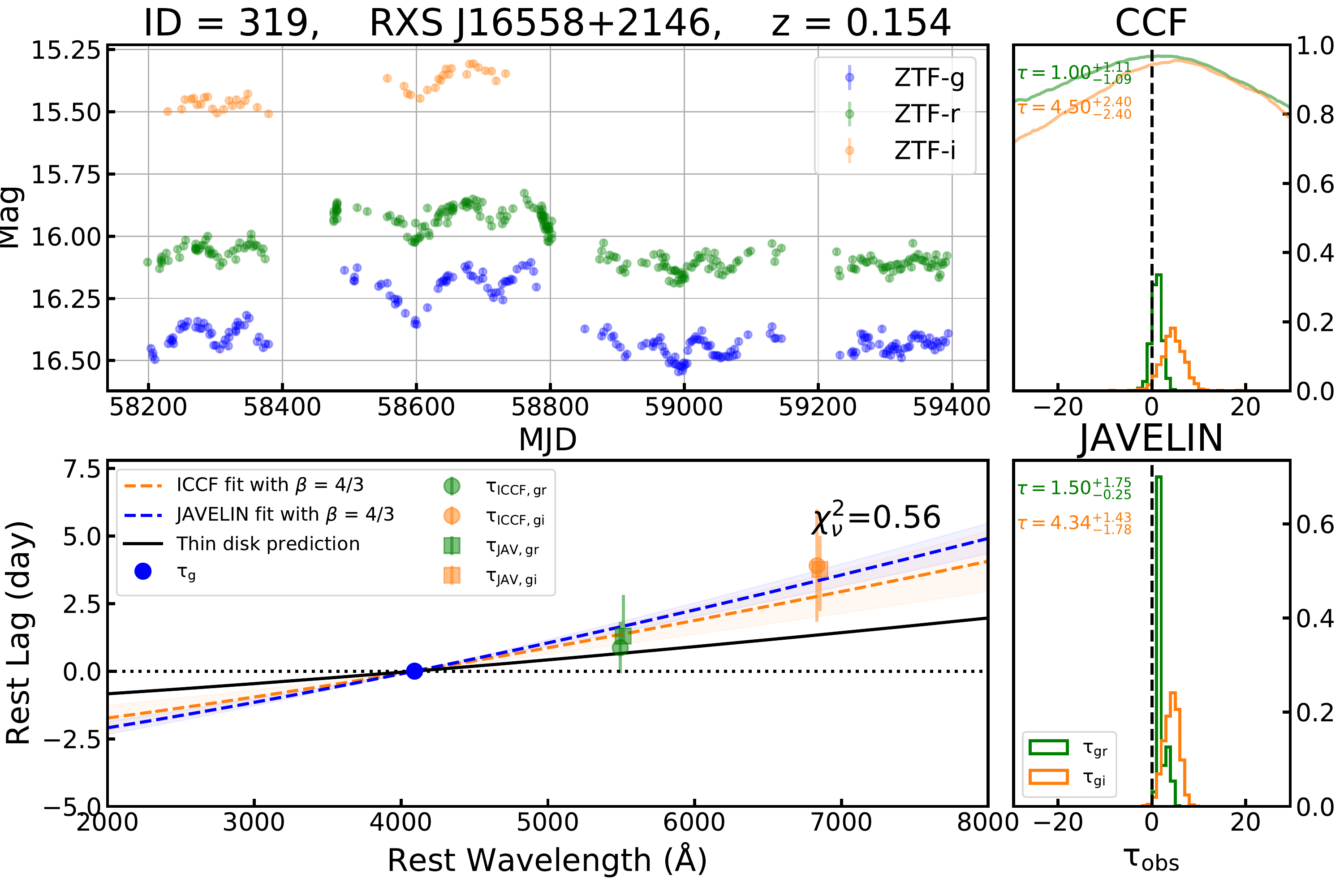} }}%
    {{\includegraphics[width=8.5cm]{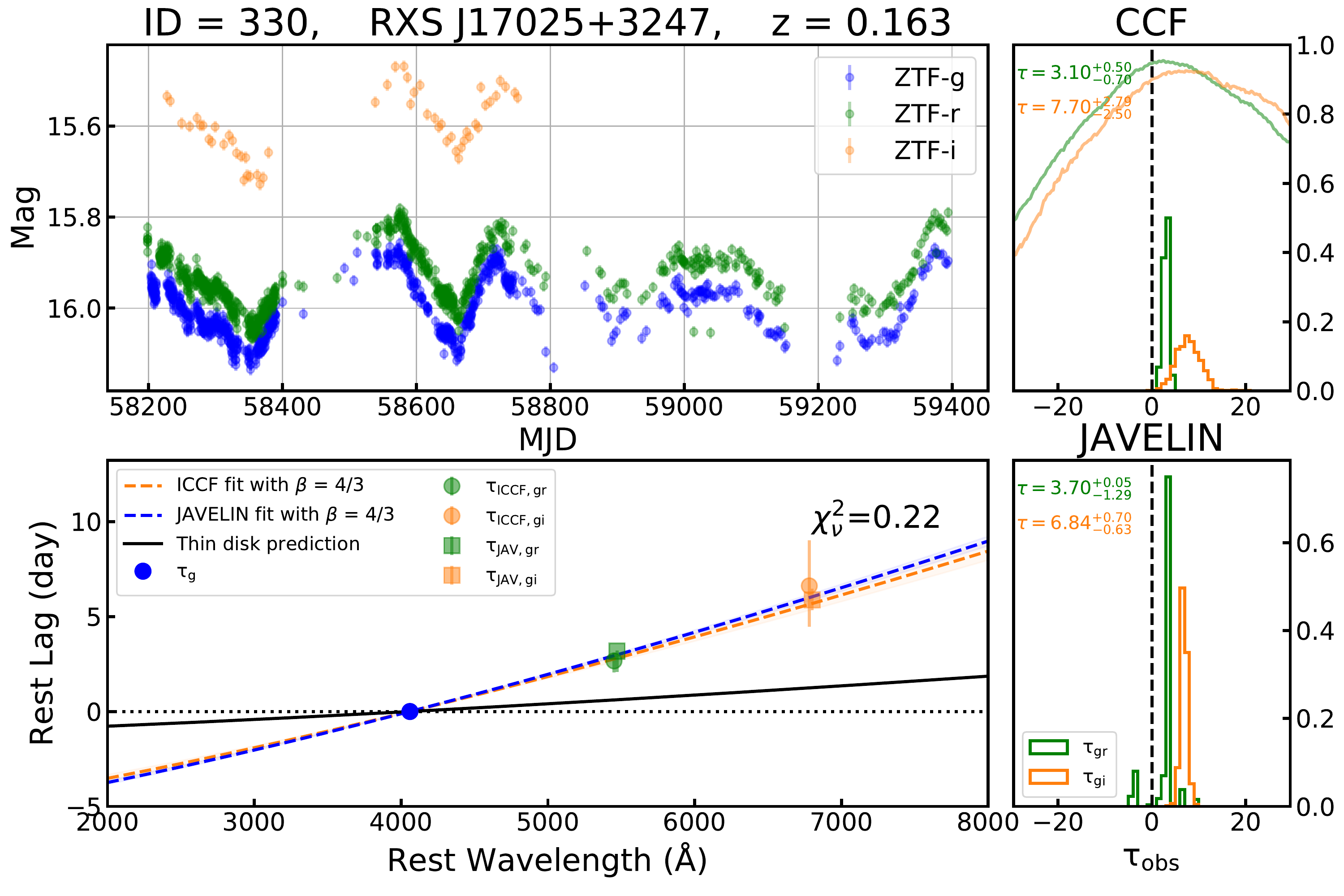} }}%
    {{\includegraphics[width=8.5cm]{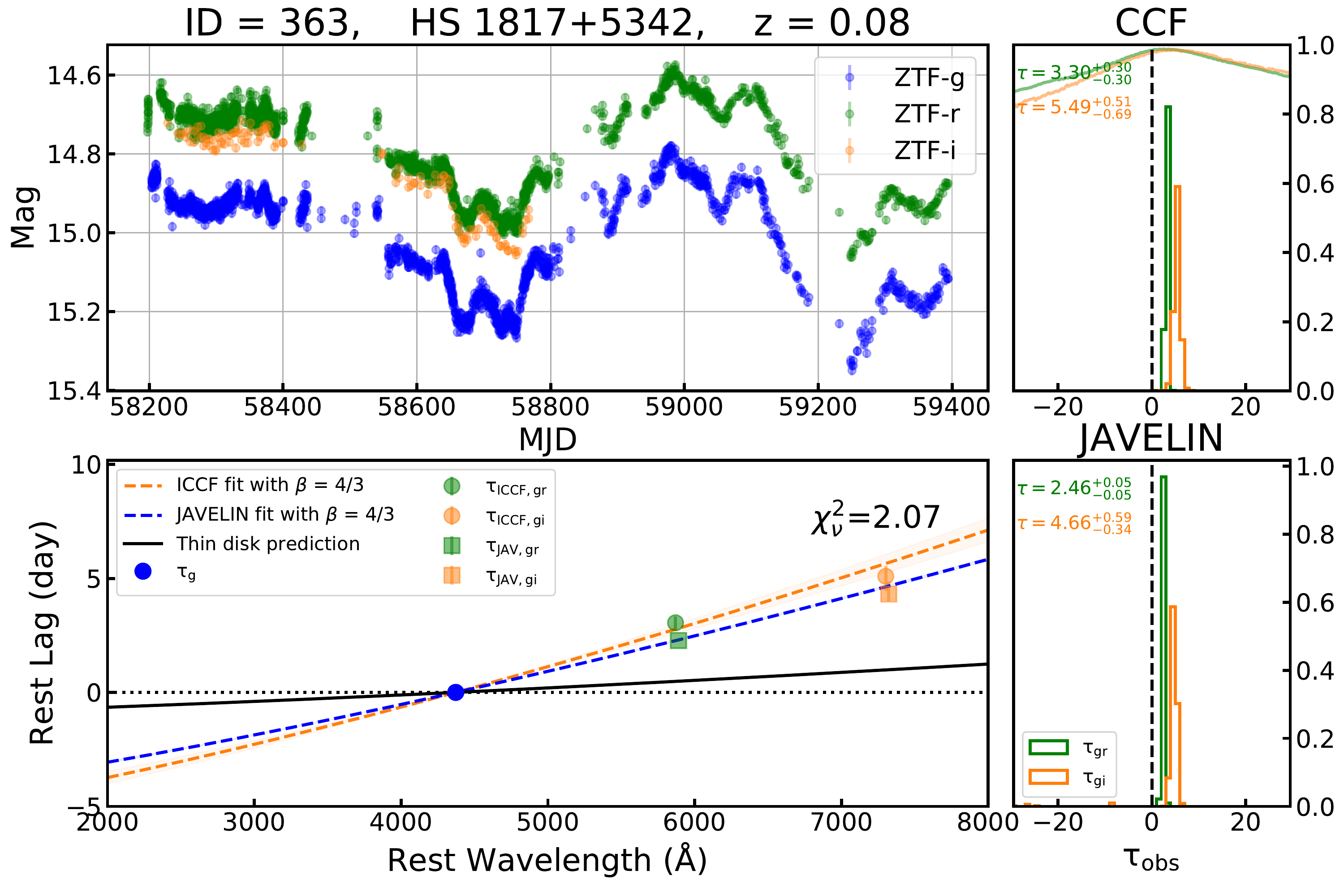} }}%
    \caption{Examples of six initial-sample AGN showing ZTF light curves, cross-correlation measurements, and wavelength-lag dependence. Upper left panel: ZTF $gri$-band light curves. Right panels: ICCF and {\tt JAVELIN} lag measurements. The upper panel shows the observed centroid lags between $g-r$ and $g-i$ bands measured from ICCF and the bottom panel presents the normalized lag posterior distributions from {\tt JAVELIN}. The lag measurements are listed. Bottom panel: comparison between the observed and SSD predicted lag-wavelength dependence. The observed lags are shifted to the rest-frame and fitted by Equation \ref{eq:tau} assuming $\beta =4/3$ and $\lambda_{0} = 4722$ \AA$/(1+z)$. The fitting results, 1$\sigma$ uncertainties of ICCF and {\tt JAVELIN}, and the reduced $\chi^2_{\nu}$ of the ICCF fits are presented. The SSD predicted wavelength-lag dependence (solid black line) is calculated with Equation \ref{eq:R}  ($\tau = R_{\mathrm{\lambda,SSD}}/c$) based on the AGN continuum luminosity and estimated BH mass. All the figures for the parent sample are available online.}   
    \label{fig:example}%
\end{figure*}

\section{Lag measurements and disk-size estimation}\label{sec:lag}
To measure reverberation lags, the most frequently used techniques are ICCF \citep[e.g.][]{Peterson98} and {\tt JAVELIN} \citep{Zu2011}. Previous investigations that studied the systematics between these two methods reached a consensus that both methods provide consistent lag measurements, and {\tt JAVELIN} usually gives relatively smaller lag uncertainties \citep{Li19,Edelson19,Yu20b}. However, {\tt JAVELIN} makes a strong assumption that the reprocessed light curve is a smoothed, shifted, and scaled version of the driving light curve, which may not always be true \citep[e.g., the BLR holiday in NGC 5548,][]{Goad16}. In this work, we adopt both methods to estimate the inter-band lags. 

\subsection{ICCF Lags and Assessment of Cross-Correlation Reliability} \label{sec:iccf}
The ICCF method estimates time lags by linearly interpolating the two light curves, shifting either light curve by a time lag, calculating the cross-correlation coefficient $r$ at this time lag, and searching for the most likely lag over a grid of values. In this work, we use the publicly available package {\tt I$^2$CCF}\footnote{\url{https://github.com/legolason/PyIICCF}} (Guo et al. in prep.) to implement ICCF lag analysis. This code implements the ICCF method \citep{Gaskell1987,White1994,Peterson98,Sun18}, and additionally provides a null hypothesis test to assess the robustness of the detected correlation. The \texttt{I$^2$CCF} code also provides the option of other non-linear interpolation methods, but for this work we employ linear interpolation following the standard and well-tested ICCF technique.

We set the lag search range in I$^2$CCF to be $\pm$30 days in the observed frame. Because our sample consists of moderate-luminosity AGN with $z <0.8$, this search range is sufficiently large with respect to the expected continuum lags \citep[usually $\lesssim$ 10 days even for some bright quasars at $z \sim$ 1, ][]{Yu20}. \cite{Li19} suggested that the overall ICCF shape does not change dramatically with different $\tau$ grid spacing (e.g., 0.1, 0.2, 0.5, 1.0, and 2.0 times the light-curve cadence), thus we adopt a 0.5-day grid spacing which is more than sufficient to sample the $\sim3$-day cadence of the ZTF light curves. 

I$^2$CCF utilizes the traditional flux randomization/random subset sampling (FR/RSS) procedure \citep{White1994,Peterson98} to obtain the lags and uncertainties. This Monte Carlo method randomizes the flux measurements by their uncertainties (FR) and randomly chooses a subset of light-curve points (RSS) to build the centroid lag distribution, whose median value and 1$\sigma$ range serve as the centroid lag and error, respectively. We carried out 1000 FR/RSS iterations for each measurement. Examples of ICCF results are displayed in Figure~\ref{fig:example}. 

We measured two different lag values: the peak and the centroid of the CCF. The centroid lags are calculated from points above $95$\% of the CCF peak value (i.e., $r > 0.95\,r_\mathrm{ max}$). This differs from the widely used $0.8\,r_\mathrm{ max}$ threshold because some of the ZTF CCFs have relatively flat peaks and decrease slowly across the search range, and the following assessment of cross-correlation reliability requires a uniform lag search range for all objects to obtain consistency across the sample. The choice of $0.8\,r_\mathrm{ max}$ would result in some objects hitting the search range boundary before reaching 80\% of $r_\mathrm{ max}$. We tested different choices of the threshold ranging from $0.8\,r_\mathrm{ max}$ to $0.95\,r_\mathrm{ max}$, and found that the lags are not sensitive to this threshold. We adopt the centroid lags as the final ICCF lag measurements, as the centroid lags are thought to more represent the luminosity-weighted radius of the reprocessing region \citep[e.g.,][]{Gaskell86}, although for these measurements the differences between the CCF peak and centroid are very small (e.g., $\tau_\mathrm{gr,cen}-\tau_\mathrm{gr,peak} = 0.2^{+0.7}_{-0.7}$ days, $\tau_\mathrm{gi,cen}-\tau_\mathrm{gi,peak} = 0.0^{+0.4}_{-0.6}$ days for the parent sample, see \S \ref{sec:parent_core}).


Next, we assess the cross-correlation reliability using I$^2$CCF. In brief, we carry out a null-hypothesis test aiming to determine the possibility of obtaining a simulated $r_\mathrm{ max, sim}$ from two random and uncorrelated light curves that is larger than the observed $r_\mathrm{ max, obs}$, under the same S/N and cadence as the observed light curves. The details will be presented in a forthcoming work and have been briefly introduced in \citet{U21}. Here we describe the basic procedures: 1) we first generate 500 mock light curves based on the damped random walk (DRW) model for each continuum band (e.g., $g$ and $r$), with the same cadence and noise level as the real observations; 2) measure the cross-correlations of the simulated $g$ and real observed $r$ band light curves (also the opposite way) to obtain the $r_\mathrm{ max}$ for each pair of simulations; 3) determine $p$-value of the null hypothesis test, which is the fraction of the simulations that achieve positive lags ($\tau>$ 0) with peak CCF values ($r_\mathrm{ max}$) equal to or higher than the observed one. This derived $p$-value thus provides an indicator of the lag reliability given the observed $r_{\mathrm{ max}}$ and light-curve properties (S/N, cadence, and duration). We emphasize that this test offers a relative indicator of cross-correlation reliability for these ZTF data, rather than a false-positive probability, because the null hypothesis of intrinsically uncorrelated data is highly implausible in normal AGN. A high $p$-value is much more likely to be an indicator of poor-quality data rather than a sign that the AGN light curves in different photometric bands might not actually be intrinsically correlated. There is no clear boundary in $p$-value between significant and insignificant correlations as it strongly depends on the design of the RM campaigns and lag search range. In this work, to distinguish different levels of lag quality according to this simulation, we use $p$-value thresholds of $<$0.1 and $<$0.2 to define the core and parent samples, respectively in \S \ref{sec:parent_core}.

\begin{figure*}
 \centering
 \includegraphics[width=0.8\textwidth]{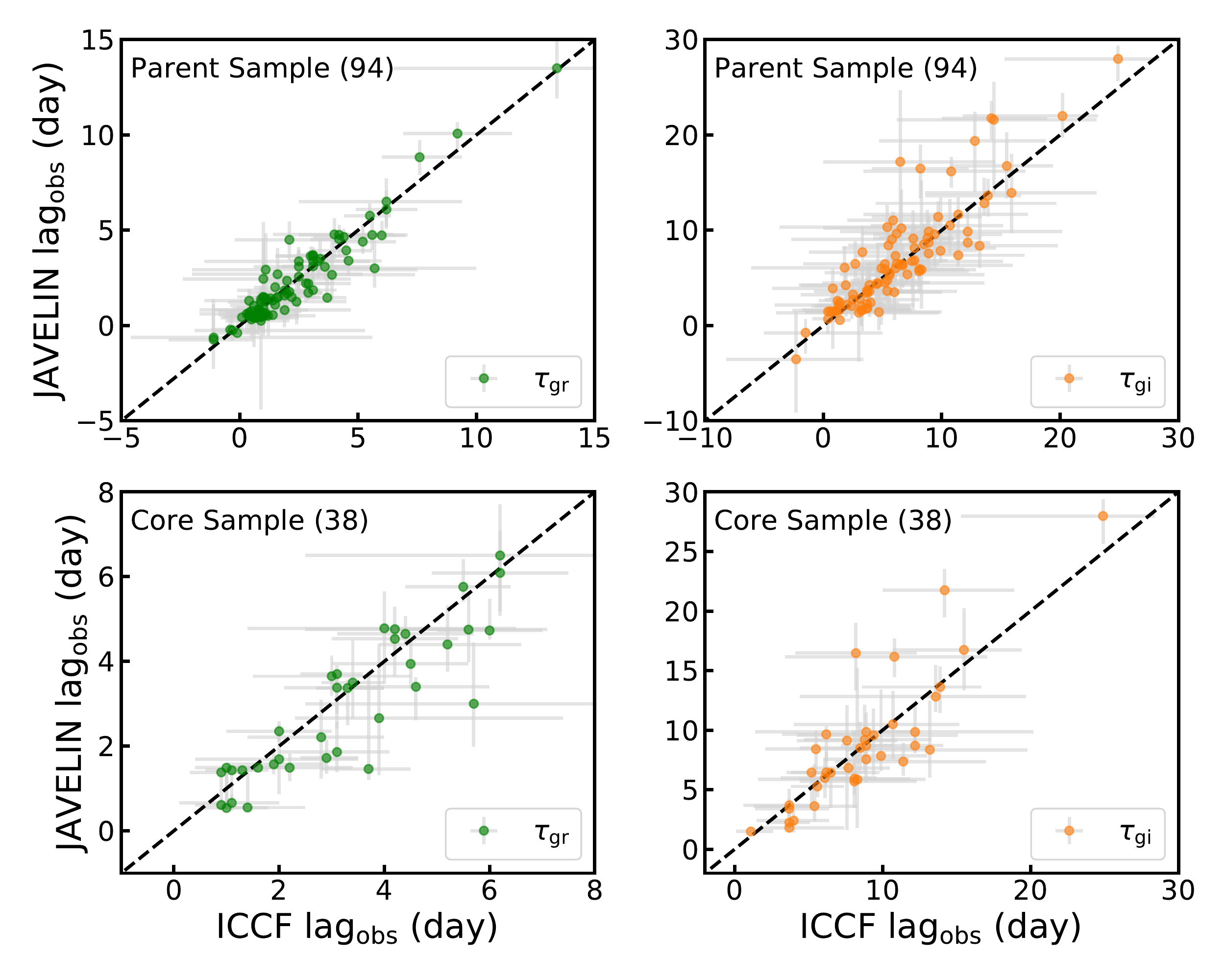}
    \caption{Observed-frame lag comparison between ICCF and {\tt JAVELIN} methods. Both $g-r$ and $g-i$ lags, their uncertainties, and sample size are shown in each panel. The 1:1 ratio is indicated by the black dashed line.}
    \label{fig:laglag}
\end{figure*}

\subsection{JAVELIN Lags}
{\tt JAVELIN} is another commonly used approach for lag measurement. Compared to the simple linear interpolation of light curves in ICCF, {\tt JAVELIN} interpolates the light curves under the assumption that the quasar statistical variability follows the DRW model \citep{Kelly09,Kelly14}, which is useful when the observations are sparse or unevenly sampled. In practice, the DRW model seems to provide a reasonable statistical description of stochastic quasar continuum variability \citep{Kelly09,Kozlowski10,Macleod10}, even though the observed power spectral slopes on very short \citep[minutes,][]{Mushotzky11,Kasliwal15} and long \citep[decades,][]{MacLeod12,Guo17} timescales deviate from the DRW predictions.

We measured the $g-r$ and $g-i$ lags separately with {\tt JAVELIN}. It first models the driving continuum ($g$-band) light curve to constrain the damping timescale and asymptotic variability, and the results serve as priors to feed the following step. The next step in {\tt JAVELIN} is modeling the driving and responding light curves with the Markov Chain Monte Carlo (MCMC) approach, assuming the flux in the $r$ and $i$ bands is a shifted, scaled, and smoothed version of the $g$-band light curve. In total, there are five free parameters in the MCMC modeling, including the damping timescale, asymptotic variability amplitude, time lag, a top-hat smoothing function, and a flux scaling factor. The lag search range is set to be the same as for the ICCF measurements ($\pm30$ days). To explore the parameter space, we used 100 walkers for each set of initial values with each walker corresponding to 30,000 steps. Burn-in phases (10,000 steps) were deleted before connecting 100 chains end to end. The final lags and their uncertainties were determined by the median value and 1$\sigma$ standard deviation of the lag posteriors. Lag posterior examples of six objects are shown in Figure \ref{fig:example}. 

The comparisons between ICCF lags and {\tt JAVELIN} lags from our two main subsamples (the sample definition will be described in \S \ref{sec:parent_core}) are shown in Figure \ref{fig:laglag}. As expected, the two methods generally produce consistent lag measurements, except for a few outliers which are mainly due to the broad distributions of the lag posteriors. The median $\tau_{\mathrm{gr}}$ ($\sim$2 days) in both samples is, as expected, smaller than that of $\tau_{\mathrm{gi}}$ ($\sim$7 days). Due to $\tau_{\mathrm{gr}}$ being smaller for most objects than the 3-day cadence of the ZTF light curves, the $g-i$ lags are usually detected more easily than $g-r$ lags, despite the fact that the $i$-band light curves are shorter and less variable. In addition, we also confirm that the ICCF lag uncertainties are larger than those from {\tt JAVELIN} by a factor of $\sim$2 in our sample. In the following analysis, we use ICCF results for statistical studies as its error estimation is more conservative and the ICCF measurements also include the cross-correlation reliability test.

\subsection{Estimation of the Disk Size}\label{sec:calsize}
Now, we attempt to convert the observed $g-r$ and $g-i$ lags to an estimate of continuum emission (CE) size at 2500 \AA, which is frequently used to compare the CE size and disk size in RM and microlensing studies. We use the term ``CE size'' ($R_\mathrm{2500,obs}$) to refer to the continuum emission size inferred by the rest-frame inter-band lag, which is likely a superposition of lag signals from different physical components, e.g., disk and BLR emission, and ``disk size'' ($R_\mathrm{ 2500,SSD}$) to refer the SSD-predicted disk size, which is dependent on BH mass and accretion rate.

To measure $R_\mathrm{ 2500,obs}$, we first transform the lags from observed frame to rest frame using the spectroscopic redshift compiled in the MQC. Then we perform a fit of Equation \ref{eq:tau} to the $g-r$ and $g-i$ lags by forcing the fit through $\tau=0$ at  reference wavelength $\lambda_0$, namely, the $g$-band wavelength 4722 \AA/(1+$z$). The 1$\sigma$ lag uncertainties are also incorporated in the fits. As suggested by previous studies, inter-band continuum lags in AGN generally follow the $\tau\sim\lambda^{4/3}$ law \citep{Fausnaugh16, Jiang17,Jha21,Yu20}. Hence we fix the power-law index ($\beta$) to 4/3, leaving $R_{\lambda_{0}}$ to be the only free parameter in the fits. The reduced $\chi^2_{\nu}$ values for one degree of freedom are listed for the example fits shown in Figure \ref{fig:example}. We also tested two-free-parameter fits with the slope $\beta$ as a free parameter; these tests are described in \S \ref{sec:color}. Finally, the best-fit value of $R_{\lambda_{0}}$ is used to derive the CE size at 2500 \AA\ and its uncertainty assuming $R_{\lambda} \propto \lambda^{4/3}$, i.e., $R_\mathrm{ 2500,obs} = R_\mathrm{ \lambda_{0}} (2500 \textrm \AA / \lambda_{0})^{4/3}$. All of the lag and disk size measurements are listed in Table \ref{tab:sample}. 

As we fixed the power-law slope of $\beta=4/3$, if the intrinsic slope of the $\tau-\lambda$ relation is far from this assumption, the fit could yield poor results and large $\chi^2_{\nu}$ values. As an example, for object ID $=$ 301 shown in Figure \ref{fig:example},  $\tau_\mathrm{gr}$ is much smaller than $\tau_\mathrm{gi}$ and has much smaller measurement uncertainty, and as a result the fitted model falls close to the $\tau_\mathrm{ gr}$ point but fails to match the observed $\tau_\mathrm{ gi}$ value. As tested, an acceptable fit to both $g-r$ and $g-i$ lags could yield a power-law slope larger than 6. There are two objects (ID 301 and ID 302) exhibiting such poor fits in the subsamples we select for further analysis, and we exclude these two objects from the slope analysis in \S \ref{sec:color}. In addition, we also avoided fitting negative continuum lags since the fit will fail if either $\tau_\mathrm{ gr}$ or $\tau_\mathrm{ gi}$ is negative ($<$ 20\% of the ICCF lags are negative in the initial sample and there are only 5 (2) objects with negative  $\tau_{\mathrm{gr}}$ ($\tau_{\mathrm{gi}}$) in the parent sample; see \S \ref{sec:parent_core})  .

We also calculate the SSD-predicted disk size at $\lambda =$ 2500 \AA\ in units of lt-day following \citet{Edelson17}:
\begin{equation}\label{eq:R}
    R_\mathrm{ \lambda,SSD} = \Bigg(X \frac{k\lambda}{hc} \Bigg)^{4/3} \Bigg[ \Bigg( \frac{GM_\mathrm{ BH}}{8\pi\sigma} \Bigg)  \Bigg( \frac{L_\mathrm{ E}}{\eta c^2}     \Bigg) (3+\kappa)\dot{m}   \Bigg]^{1/3},
\end{equation}
where $G$ is the gravitational constant, $M_\mathrm{ BH}$ is the \hb-based BH mass, $\sigma$ is the Stefan-Boltzmann constant, $L_\mathrm{ E}$ is the Eddington luminosity, $\dot{m}$ is the Eddington ratio,  $\eta $ is the radiative efficiency, and $\kappa$ represents the relative contribution between X-rays and viscosity in disk heating. In the calculation of bolometric luminosity and Eddington ratio, we use the 5100~\AA\ luminosity and bolometric correction factor of 9.26 \citep{Richards06}. The factor $\eta $ is assumed to be 0.1, and $\kappa$ is fixed to 1, which means an equal contribution between X-rays and accretion-disk viscosity. The factor $X$ is chosen to be 2.49, which corresponds to emission from a flux-weighted radius at a given wavelength. Later, we will also discuss the influence of using a responsivity-weighted (rather than flux-weighted) radius, corresponding to $X = 3.36$ \citep[e.g.,][]{Tie18}. The predicted $R_\mathrm{ 2500,SSD}$ is listed in Table \ref{tab:sample} and the corresponding lag can be easily obtained from $\tau = R_{\mathrm{\lambda,SSD}}/c$.

\subsection{Parent and Core Sample}\label{sec:parent_core}
Based on the outcome of the lag measurements (i.e., the lag quality assessment, lag consistency between measurement methods, and uncertainties), BH mass, and radio flux density, we conduct a further sample selection and identify two major subsamples from the initial sample of 455 objects: a parent sample with reliable lag measurements and a core sample with the best-quality lags to balance the sample size and lag quality in the following exploration of different relationships. The selection criteria for these subsamples and the numbers of objects remaining after each step are listed below. \\

\textbf{Parent sample selection:}
\begin{enumerate}
    \item $\tau+\sigma_\mathrm{ up}> 0$ for both bands and both lag measurement methods (369);
    \item All uncertainties $\sigma < 10 $ days (232);
    \item Lag ratio $ 0.2 < \tau_\mathrm{ ICCF}/\tau_\mathrm{ JAV} < 5$ (153);
    \item Lag difference $ |\tau_\mathrm{ ICCF} - \tau_\mathrm{ JAV}| < 3\sigma_\mathrm{ average}$ (112);
    \item Cross-correlation reliability $p$-value $<$ 0.2 (111);
    \item Radio flux density $F_\mathrm{ 1.4G} < 100$ mJy at 1.4 GHz  (110);
    \item Known $M_\mathrm{ BH}$ based on \hb\ (94).
\end{enumerate}
This yields 94 objects in the parent sample. Criteria 1 and 2 mainly rule out negative lags, and objects having poorly constrained lag posteriors, i.e., a very broad lag distribution or lag distribution strongly affected by aliasing peaks. Criteria 3 and 4 enforce lag consistency between ICCF and {\tt JAVELIN} both in ratio and  difference. Criterion 5 requires the $p$-value to be $<$0.2, which means two uncorrelated red-noise light curves with the same observational conditions (e.g., cadence, error, and duration) have a $<$20\% possibility to achieve an equal or higher $r_\mathrm{ max}$ than the observed value over the positive lag search range. Almost all objects passing step 4 also satisfy this criterion due to the high cadence of the ZTF light curves. Objects with strong radio emission are rare in our sample, and criterion 6 requires the radio flux density to be less than 100 mJy in order to exclude some blazars, whose continuum emission in optical can be dominated by the jet contribution (synchrotron radiation). Finally, the BH mass is needed to estimate the SSD disk size.   \\

\textbf{Core sample selection:}
\begin{enumerate}
    \item $\tau-\sigma_\mathrm{ low}> 0$ for both bands and both lag measurement methods (93);
    \item All uncertainties $\sigma < 10 $ day (81);
    \item Lag ratio $ \frac{1}{3} < \tau_\mathrm{ ICCF}/\tau_\mathrm{ JAV} < 3$ (72);
    \item Lag difference $  |\tau_\mathrm{ ICCF} - \tau_\mathrm{ JAV}| < 3\sigma_\mathrm{ average}$ (49);
    \item Cross-correlation $p$-value $<$ 0.1 (48);
    \item Radio flux density $F_\mathrm{ 1.4G} < 100$ mJy at 1.4 GHz (48);
    \item Known $M_\mathrm{ BH}$ based on \hb\ (38);
\end{enumerate}
This more stringent selection results in 38 objects in the core sample. All of these objects also belong to the parent sample. The inter-band lags are restricted to be larger than zero within 1$\sigma$ uncertainty, rejecting a large fraction of objects with very small lags unresolved by the ZTF light curves, especially for $g-r$ lags. Lag consistency is refined in criterion 3 to further eliminate some lags affected by a secondary peak or an asymmetric lag posterior. Also, criterion 5 further tightens the cross-correlation reliability. The lags for the parent and core samples are demonstrated in Figure \ref{fig:laglag}.

The main differences between the two sample selections involve the choice of how strictly to exclude lag measurements that are consistent with zero (criterion 1) and the degree of lag consistency between different methods (criterion 3). The parent sample includes some objects with  $\tau_{\mathrm{gi}}$ significantly above zero but with $\tau_{\mathrm{gr}}$ consistent with zero within uncertainties. This is because some short-timescale variability features are not fully resolved by the $\sim$3-day cadence of the ZTF data. If these objects are not considered, the sample selection can be biased toward finding longer lags, thus overestimating the disk-size discrepancy between observations and SSD predictions \citep{Homayouni19}. Therefore, we established the two samples with or without very short lags to test for this potential bias. In the latter sample without many short lags, we have a more stringent cut on lag ratio, which will also primarily affect the short lags as the longer ones are not significantly different between the two methods considering the limited lag search range. This criterion also rejects some inconsistent short lags measured by ICCF and {\tt JAVELIN}. In addition, both samples include some objects (15 for the parent and 2 for the core sample) with $\tau_{\mathrm{gi}} < \tau_{\mathrm{gr}}$ either from ICCF or {\tt JAVELIN}. These results are unexpected in the context of the lamp-post model, but they are still reasonable considering the lag measurement uncertainties, the influence of Balmer and Paschen jumps from diffuse continuum emission \citep{Cackett18,Guo22}, or other potential disk models \citep[e.g.,][]{Cai18,Sun20a}.

Comparing our sample to that of another recently published ZTF disk RM study by \citet{Guowj22}, the sample sizes of objects having significant lags are similar ($\sim 90$), and the AGN redshifts are restricted to $\lesssim$ 0.8 in both studies. However, the selected AGN are quite different: their objects are all collected from the SDSS DR14Q (which includes BH mass estimates for all objects) while our AGN are collected from the MQC with a larger sample size but without existing BH masses for some objects. By cross-matching our parent (core) sample against that of \citet{Guowj22}, we found only 4 (1) objects overlapping between the two studies. In addition, our sample selection criteria are more stringent in terms of our requirements on lag consistency and uncertainties. Our $g-r$ lags are based on the four-year DR7 light curves, while \citet{Guowj22} used DR3 light curves restricted to a one-year duration having good sampling. Although there are similarities in the scientific goals between their study and ours, our sample and scientific focus are considerably different and the two studies are complementary. 

We also checked the overlap between our sample and the sample of nearby reverberation-mapped AGN from \citet{BentzKatz15} since future continuum reverberation mapping targets are likely selected from this catalog. By cross-matching with this catalog of 86 AGN, we found that 19 of them have ZTF continuum lag measurements in our initial sample. Only one of 19 objects, Mrk~1383, is in our parent sample but not the core sample. Overall, the $g-r$ lags of these 19 AGN are in general well constrained and only a few objects show negative lags or have lags consistent with zero. However, most of the $g-i$ lags have large uncertainties due to the limited baseline of $i$-band light curves. 

Among these 19 objects, two AGN (NGC 4151 and Mrk 817) were also targets of recent intensive continuum RM programs. For NGC 4151 (ID = 98), no conclusive lags can be obtained from the ZTF data as the scatter in the $g$- and $r$-band light curves is relatively large. According to \citet{Edelson17}, the lag between 2000 \AA\ and 5000 \AA\ in NGC 4151 is less than 1 day, which would be difficult to resolve with the 3-day ZTF cadence. As for Mrk 817 (ID = 204), the observed centroid lags ($\tau_{\mathrm{gr,ICCF}} = 1.8^{+0.6}_{-0.8}$ days and $\tau_{\mathrm{gi,ICCF}} = 4.9^{+1.8}_{-2.5}$ days) are consistent with the results from the AGN STORM 2 project within uncertainties ($\tau_{\mathrm{gr}} = 1.5^{+0.9}_{-0.8}$ days and $\tau_{\mathrm{gi}} = 2.3^{+0.7}_{-0.9}$ days) \citep{Kara21}. However, this object is still rejected by our parent sample selection criteria as it does not satisfy the lag difference criterion (criterion 4).

\section{Results and Discussion}\label{sec:results}
In this section, we compare our observational results with SSD predictions in the framework of the lamp-post model.  
\subsection{Lag-Wavelength Dependence } \label{sec:color}
A key result from previous disk RM campaigns of local AGN is that there is indeed a positive correlation between inter-band lags and wavelength across $\sim$1000 to 10000 \AA, and the power-law slope of the $\tau-\lambda$ dependence is broadly consistent with $\beta=4/$3, as predicted by the SSD model. However, it is possible that some disks are in other accretion states, e.g., a slim disk, which will give a $\tau\propto \lambda^{2}$ relation \citep{Cackett20}. 

\begin{figure}
 \centering
 \includegraphics[width=0.5\textwidth]{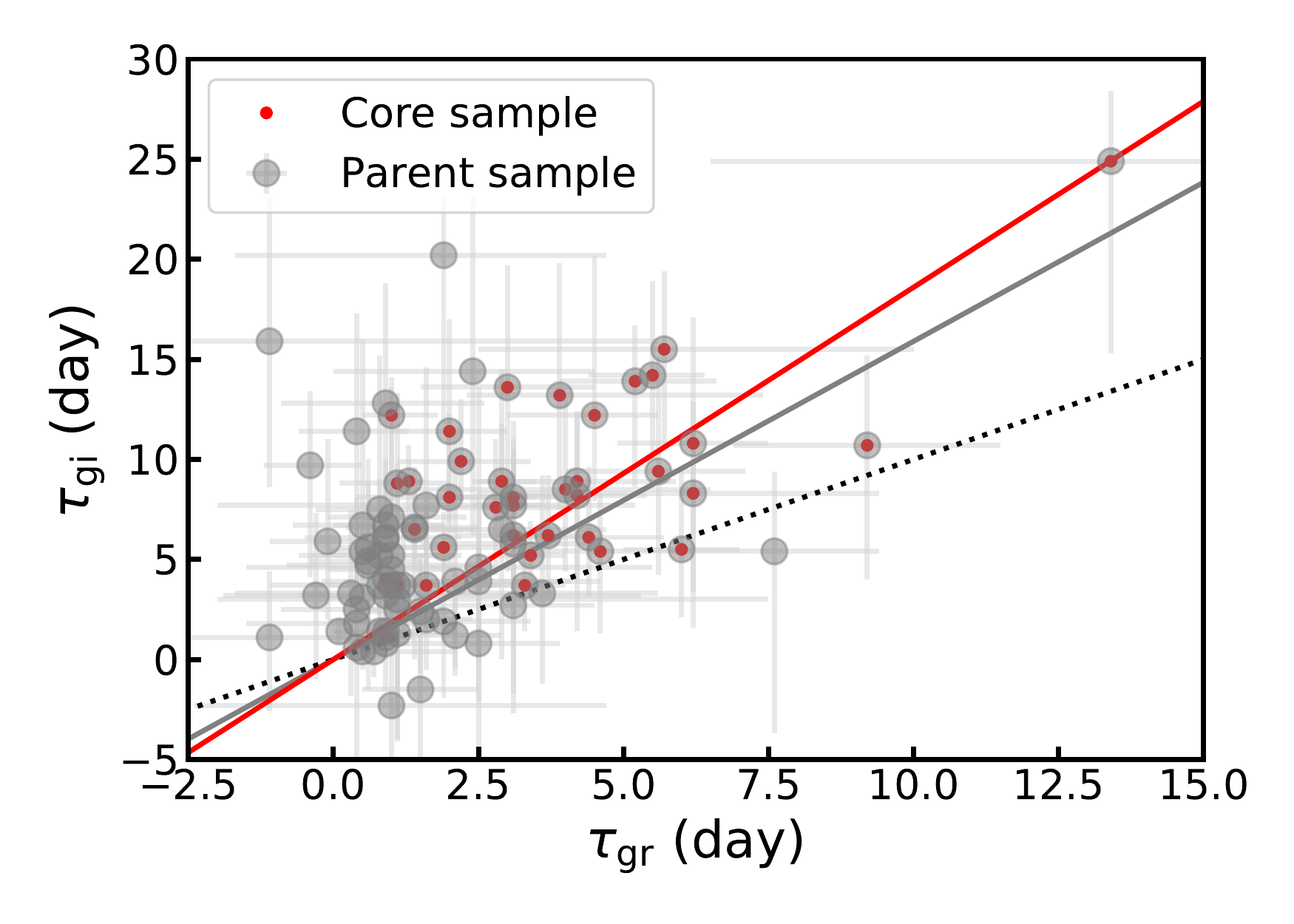}
    \caption{ICCF rest-frame lag measurements for the parent and core samples. The grey and red dots are inter-band lags with 1$\sigma$ uncertainties in parent and core sample, respectively. The grey and red lines are corresponding best linear fits through zero.  The 1:1 ratio is marked with the black dotted line.}
    \label{fig:gr-gi}
\end{figure}

As shown in Figure \ref{fig:gr-gi}, $\tau_{\mathrm{gi}}$ is usually larger than $\tau_{\mathrm{gr}}$, with confidence levels of $\sim$1.7$\sigma$ and $\sim$2.3$\sigma$ (based on the median lags and uncertainties) for the parent and core sample, respectively. Recalling that we do not force $\tau_{\mathrm{gi}}$ to be larger than $\tau_{\mathrm{gr}}$ in the sample selection, this result confirms that the inter-band lag generally rises with increasing wavelength, which is in good agreement with previous results. In a few objects we find that $\tau_{\mathrm{gi}}$ is shorter than $\tau_{\mathrm{gr}}$, which could be due to either measurement uncertainty or a physical mechanism that differs from the simple disk reprocessing model. We present one example in which the $r$-band light curve temporally leads the $g$-band in Appendix \ref{app:abnormal}, which may imply a different variability mechanism dominating over a short timescale in this object.   

\begin{figure}
 \centering
 \includegraphics[width=0.5\textwidth]{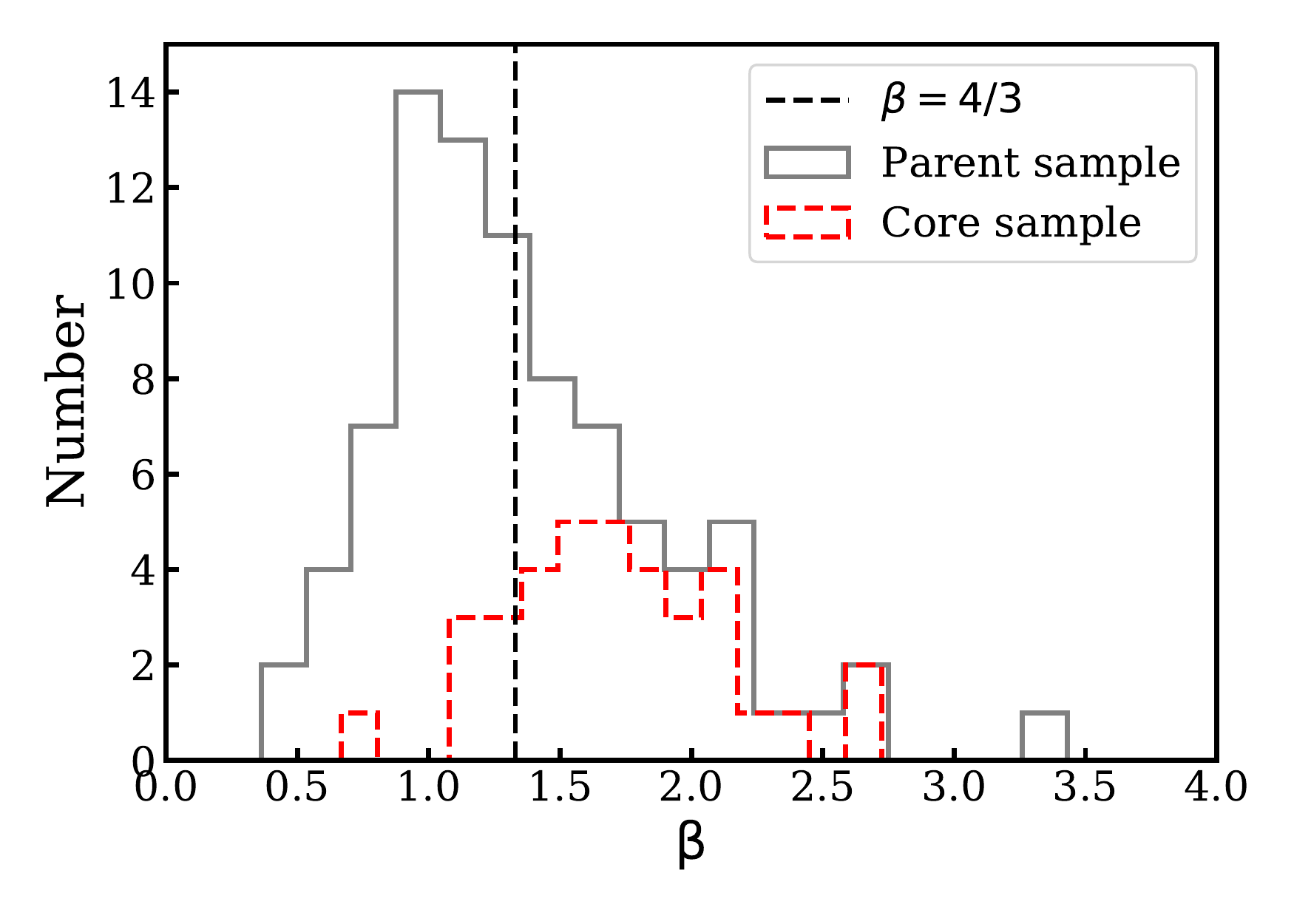}
    \caption{Distributions of $\beta$ in the parent and core samples. The SSD prediction of $\beta=4/3$ is indicated by the black dashed line. }
    \label{fig:slope}
\end{figure}

Although the wavelength range and cadence of the ZTF data only provide limited information on the value of the power-law slope $\beta$ of the lag-wavelength relationship, our measurements do allow some tests to be performed. According to Equation \ref{eq:tau}, the lag ratio between $g-i$ and $g-r$ is expected to be:
\begin{equation}
    \rm \frac{\tau_{gi}}{\tau_{gr}} = \frac{(\lambda_{i}/\lambda_{g})^{4/3}-1}{(\lambda_{r}/\lambda_{g})^{4/3}-1},
    \label{eq:lagratio}
\end{equation}
where the effective wavelengths the of ZTF $gri$ bands are 4722 \AA, 6339 \AA, and 7886 \AA. This results in a theoretical ratio $\tau_{\mathrm{gi}}/\tau_{\mathrm{gr}} = 2.0$. Otherwise, for AGN containing a slim accretion with $\tau \propto \lambda^{2}$, the theoretical ratio should be 2.2.

To estimate the slope of the $\tau_\mathrm{ gi}-\tau_\mathrm{ gr}$ relation, we first performed a linear fit ($y = \alpha x$) with the {\tt Linmix} code \citep{Brandon07} accounting for uncertainties on both quantities, based on a Bayesian model through MCMC analysis. The obtained best-fit linear relations are $\tau_\mathrm{ gi}/\tau_\mathrm{ gr} = 1.59_{-0.05}^{+0.11}$ for the parent sample and $\tau_\mathrm{ gi}/\tau_\mathrm{ gr} = 1.86_{-0.05}^{+0.14}$ for the core sample, respectively. These results do not change significantly if we drop one obvious outlier which shows the longest lags in Figure \ref{fig:gr-gi}. We emphasize that the scatter in the $\tau_{\mathrm{gr}}- \tau_{\mathrm{gi}}$ relation is quite large, and the data do not obey a simple proportionality between the two quantities as predicted by Equation \ref{eq:lagratio}.

Additionally, we carried out power-law fits using Equation \ref{eq:tau} to the observed lags, allowing both $R_{\lambda_{0}}$ and $\beta$ to be free parameters. Figure \ref{fig:slope} shows the distributions of power-law index $\beta$ for the parent and core samples, which have $\beta = 1.1^{+0.7}_{-0.4}$ and $1.6^{+0.4}_{-0.5}$ (median value and 1$\sigma$ dispersion) respectively. If we further exclude objects in the redshift range from $z =$ 0.1 to 0.3, where \ha\ falls in the $i$-band filter, the median value of $\beta$ will be slightly reduced by 0.1 to 0.2.  

Both tests indicate that the inter-band lags increase with wavelength in core sample, generally consistent with the SSD predictions. However, the slope is relatively flatter in the parent sample with respect to that of the core sample, which is likely to be caused by the less reliable lag measurements: the shorter lags are usually less reliable, and these small lags could yield a flatter slope. Otherwise, it may indicate that an extra component may play a role in flattening the slope of the $\tau-\lambda$ relation. 

The presence of broad emission lines and diffuse BLR continuum emission in the ZTF filter bands will affect the lag measurements and thus the power-law slope of the $\tau-\lambda$ relation. For AGN at redshift $z \lesssim 0.8$, the simulation presented in \citet{Guowj22} (see their Figure 14) suggests that BLR components of different emission lines overall tend to shorten the $g-r$ and $g-i$ lags by $\sim$0.6 day compared with the intrinsic continuum lags, since the $g$ band usually includes more broad emission-line flux although the prominent \ha\ line is usually in the $r$ or $i$ band for low-redshift AGN. This implies that our $\tau_\mathrm{ gr}$ and $\tau_\mathrm{ gi}$ values could be $\sim$40\% (19\%) and 11\% (7\%) (based on median lags) smaller than the values without contamination from BLR emission lines for the parent (core) sample. Thus the emission-line contribution tends to flatten the slope of the $\tau-\lambda$ relation in our sample.

\begin{figure}
 \centering
 \includegraphics[width=0.5\textwidth]{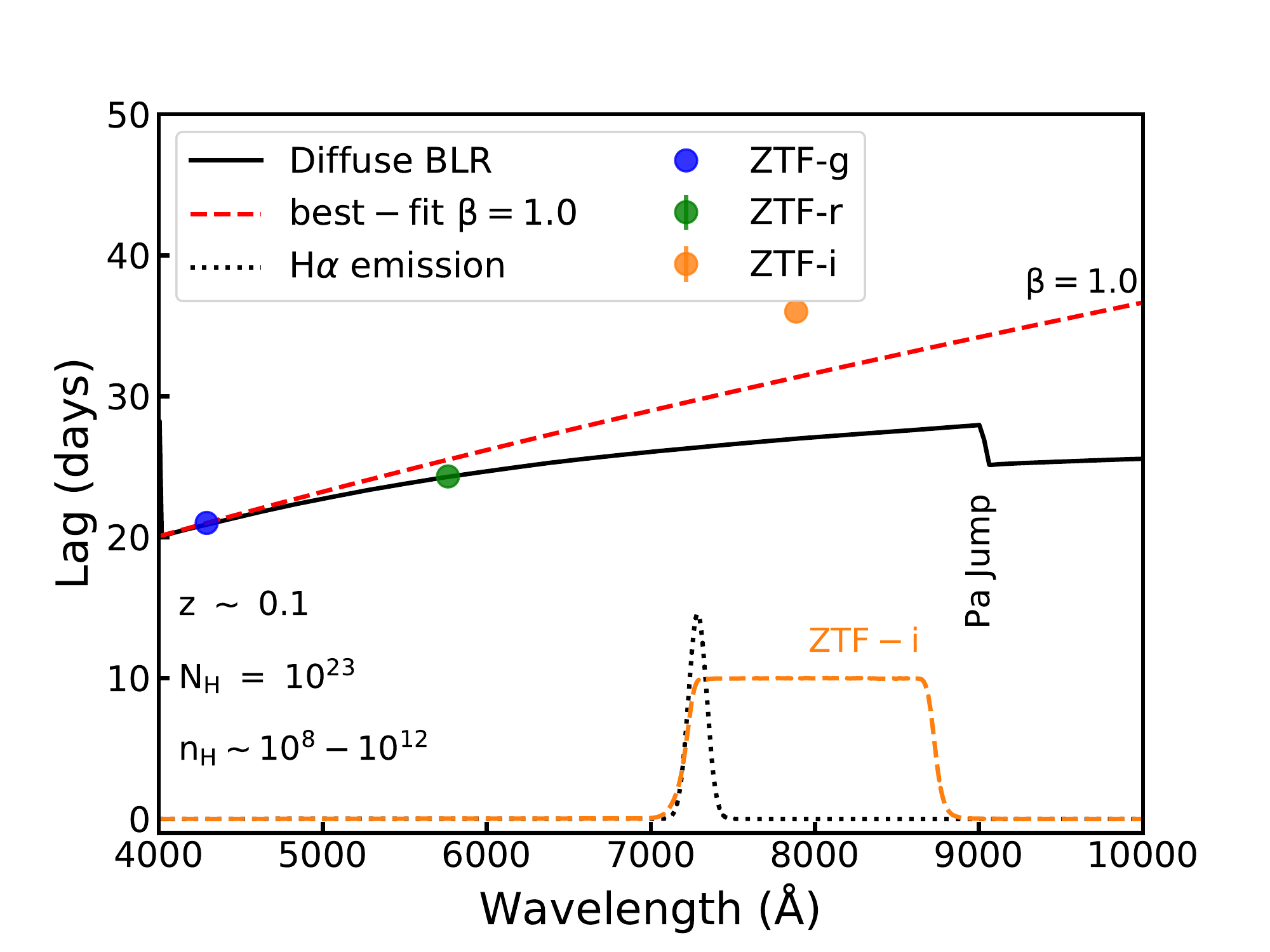}
    \caption{Estimate of the slope of the lag-wavelength relation across the wavelength range of ZTF filters for BLR emission, including diffuse continuum emission and an \ha\ line with a FWHM of 6000 \kms. The \ha\ lag is assumed to be 10 days. The lag spectrum of diffuse BLR emission is from \citet{Korista19}  with hydrogen column density $\log(N_\mathrm{H}/\mathrm{cm}^{-2}) = 23$ and hydrogen gas density $\log(n_\mathrm{H}/\mathrm{cm}^{-3})$ integrated from 8 to 12 dex. A typical redshift of $z\sim $ 0.1 is assumed. The orange dashed line indicates the ZTF $i$-band transmission function. Three lag values at the effective wavelengths of ZTF filters are fitted with Equation \ref{eq:tau}.}
    \label{fig:dc}
\end{figure}

On the other hand, the diffuse BLR continuum emission, dominated by free-free and free-bound emission from BLR clouds, is thought to make a major contribution to inter-band lags at rest-frame optical wavelengths. Figure \ref{fig:dc} presents an approximate estimation of the slope $\beta$ of the $\tau-\lambda$ relation assuming that the continuum lags are solely due to a responsivity-weighted lag spectrum of diffuse BLR continuum emission \citep{Korista19} and a broad \ha\ lag. The emitting size ($R_{\lambda_{0}}$) of the $g$ band is around 21 lt-days according to the lag spectrum of diffuse BLR emission and the \ha\ lag contribution to the $i$ band in a typical AGN ($M_\mathrm{ BH}$ $\sim$ $10^8~M_{\odot}$ at $z\sim0.1$) is $\sim$10 days assuming that 20\% of the $i$-band flux is from \ha. An identical power-law fit with Equation \ref{eq:tau} is applied to the data points which are the combined lags of diffuse BLR and \ha\ emission at effective wavelengths of the ZTF filters. The best-fit slope is 1.0 with an uncertainty of 0.1 considering different BLR environments (e.g., column density). If only a pure diffuse BLR is considered, the slope is $\sim$ 0.5, much flatter than the SSD prediction. This indicates that if the continuum lag is contaminated by the diffuse BLR contribution, we should expect a slope flatter than 4/3, yet steeper than 0.5. 

Overall, both BLR emission lines and diffuse BLR emission could influence the observed lags and the slope of the $\tau-\lambda$ relation in our sample, with different impact on different objects depending on redshift and spectral shape. More dedicated monitoring across a broader range of wavelengths is needed in the future to further explore this problem, along with modeling of the AGN spectra to constrain the contributions of different line and continuum components to each filter.

\begin{deluxetable*}{lcccccccl}[htbp]
\caption{Properties of 12 continuum-mapped AGN} \label{tab:localAGN}
\tablewidth{0pt}
\tablehead{
\colhead{Name} &
\colhead{$z$} &
\colhead{$R_\mathrm{ 2500}$}&
\colhead{$R_\mathrm{ 2500,err}$}&
\colhead{$R_\mathrm{ 2500,SSD}$}&
\colhead{log $L_\mathrm{ 5100}$}&
\colhead{log $M_\mathrm{ BH}$}&
\colhead{$\dot{m}$}&
\colhead{Reference}\\
\colhead{} &
\colhead{} &
\colhead{(lt-day)}&
\colhead{(lt-day)}&
\colhead{(lt-day)}&
\colhead{(erg s$^{-1}$)}&
\colhead{($M_\mathrm{ \odot}$)}&
\colhead{} 
}
\startdata
 Ark 120        &0.033 &   0.96 &       0.22 &0.62&  44.31 &  8.18 &0.10& \citet{Lobban20} \\
 Fairall 9      &0.047 &   1.68 &       0.09 &0.71&  44.25 &  8.41 &0.05& \citet{HernandezSantisteban20}\\
 MCG +08-11-011 &0.021 &   0.44 &       0.04 &0.16&  43.31 &  7.45 &0.05& \citet{Fausnaugh18}\\
 Mrk 110        &0.035 &   0.34 &       0.04 &0.18&  42.63 &  7.29 &0.15& \citet{Vincentelli21}\\
 Mrk 142        &0.045 &   0.45 &       0.02 &0.06&  43.29 &  6.23 &0.86& \citet{Cackett20}\\
 Mrk 509        &0.026 &   1.14 &       0.14 &0.55&  44.28 &  8.05 &0.13& \citet{Edelson19}\\
 Mrk 817        &0.031 &   1.26 &       0.11 &0.27&  42.82 &  7.59 &0.13& \citet{Kara21}\\
 NGC 2617       &0.014 &   0.21 &       0.04 &0.10&  42.63 &  7.51 &0.01& \citet{Fausnaugh18}\\
 NGC 4151       &0.003 &   0.49 &       0.16 &0.12&  42.73 &  7.60 &0.01& \citet{Edelson17}\\
 NGC 4593       &0.009 &   0.20 &       0.03 &0.05&  42.37 &  6.88 &0.02& \citet{Cackett18}\\
 NGC 5548       &0.017 &   0.82 &       0.01 &0.24&  43.51 &  7.72 &0.05& \citet{Fausnaugh16}\\
 PG 2308+098    &0.433 &   9.46 &       1.71 &5.78&  45.83 &  9.60 &0.12& \citet{Kokubo18}\\
\enddata
\tablecomments{$R_\mathrm{ 2500}$ is calculated based on the $R_\mathrm{ \lambda_{0}}$ from original literature with Equation \ref{eq:tau} assuming $\beta=4/3$, and the errors are propagated from the error of $R_\mathrm{ \lambda_{0}}$. All of them are nearby AGN except PG 2308+098 at $z = 0.433$. }
\end{deluxetable*}

\begin{figure*}
 \centering
 \includegraphics[width=1\textwidth]{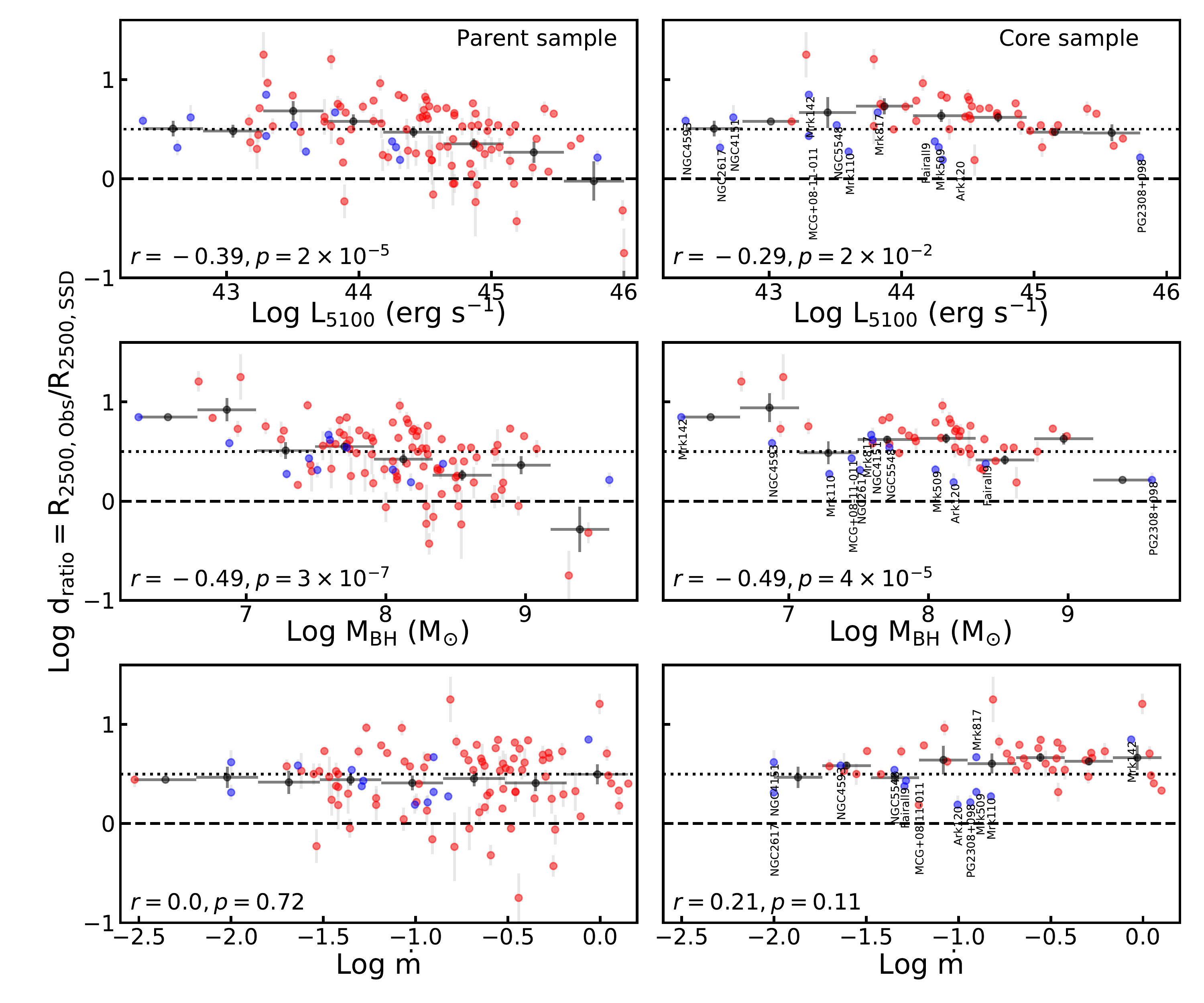}
    \caption{Ratio between continuum emission size from observation and SSD prediction as a function of basic AGN properties. In each panel, the binned mean values and 1$\sigma$ uncertainties are plotted for better demonstration of the correlations. Red dots are ZTF parent/core-sample objects and blue dots are local AGN from Table \ref{tab:localAGN}. The Spearman correlation results for the P$+$L and C$+$L samples are listed in each panel. Black dashed and dotted lines indicate $d_\mathrm{ ratio} = $ 1 and 3, respectively.}
    \label{fig:disk_ratio}
\end{figure*}

\subsection{Larger-than-expected Disk Size}\label{sec:disksize}
In previous disk RM campaigns of individual nearby type 1 AGN, the CE sizes are usually larger than the expected disk sizes \citep[e.g.,][]{Fausnaugh16,Fausnaugh18,Edelson19}. On the other hand, studies of more luminous and distant quasars from large sky surveys claimed more comparable size between them \citep[e.g.,][]{Homayouni19,Yu20}. This might be caused by poorer data quality, for example, lower cadence and/or lower accuracy of the observed light curves from large sky surveys relative to dedicated, intensive RM campaigns of nearby AGN. Alternatively, this difference could be intrinsic and caused by some underlying physics \citep[e.g., an intrinsic $L-\tau$ dependence,][]{Li21}. 

Before examining the disk size discrepancy for the ZTF sample, we collected data for several nearby AGN from individual dedicated disk RM campaigns (we refer to these as the ``local sample'') to extend the parameter space of our study in terms of luminosity and BH mass coverage. Their properties are listed in Table \ref{tab:localAGN}. The disk size predicted by the SSD model at 2500 \AA\ is calculated following \S \ref{sec:calsize} based on lag measurements from the literature. Their 1$\sigma$ uncertainties are derived from fitting the rest-frame inter-band lags with Equation~\ref{eq:tau}. We have excluded lags between the $u/U$ bands and other bands due to a potentially strong contamination from the Balmer continuum. 

Figure \ref{fig:disk_ratio} displays the correlations between size ratio ($d_\mathrm{ ratio}=R_\mathrm{ 2500,obs}/R_\mathrm{ 2500,SSD}$) and luminosity, BH mass, and Eddington ratio, respectively, for the ZTF samples and the local sample. Two features are apparent: the CE size indicated from observation is typically a factor of $\sim$3 larger than the SSD prediction, and the size ratio anti-correlates with continuum luminosity and with BH mass. The median $d_\mathrm{ ratio}$ values are 3.1 and 3.7 (with a scatter of 0.4 dex) for the parent+local (P$+$L) and core$+$local (C$+$L) samples, respectively. The median ratios are 3.1, 2.5, and 4.0, when calculated separately for the parent, core, and local samples. Our result confirms the discrepancy found in previous disk RM studies of individual low-redshift AGN that the observed CE size is a factor of $\sim$3 larger than SSD disk predictions when $X$ = 2.49 is assumed. Compared to previous disk studies, our work increases the sample size to $\sim$100 AGN.

Alternatively, if $X$ = 3.36 is used instead of 2.49 in the disk size calculation to account for systematic uncertainties in the conversion of temperature to wavelength at a given radius \citep{Tie18}, the predicted SSD disk radii would be larger by a factor of $(3.36/2.49)^{4/3}=1.5$. This would result in a reduced size discrepancy with $d_\mathrm{ratio}\approx2$, consistent with the results from \citet{Guowj22}. For previous studies, \citet{Yu20} pointed out that the CE sizes in most of their objects are larger than SSD prediction if the adopted $X=3.36$ was switched to $X=2.49$, which is consistent with our findings. However, best-fit disk sizes with $X=2.49$ are still consistent with the SSD prediction within 1.5$\sigma$ in the sample of \citet{Homayouni19}.

Furthermore, Figure \ref{fig:disk_ratio} also confirms that $d_\mathrm{ ratio}$ is anticorrelated with both continuum luminosity and black hole mass, as first reported by \citet{Li21} based on a sample of 10 nearby AGN plus measurements from \citet{Homayouni19} and \citet{Yu20}. Both the Spearman correlation analysis results and binned trends are shown in Figure \ref{fig:disk_ratio}, indicating strong anti-correlations (except for a mild one with luminosity in the C$+$L sample\footnote{The anti-correlations with luminosity are clear when respectively considering the core or local samples. However, it becomes ambiguous when two samples are combined. This may be caused by the limited sample size or different data quality in ZTF relative to local AGN. We still consider the mild anti-correlation from the core sample as our fiducial one due to the more reliable lag detections.}). In our sample, the correlation between $d_\mathrm{ ratio}$ and BH mass is stronger than that between $d_\mathrm{ ratio}$ and luminosity, in contrast to the results from \citet{Li21}. This does not necessarily indicate that BH mass is a more fundamental driving factor than luminosity, considering the limited sample size and the large uncertainties. As for Eddington ratio, there is little correlation between $d_\mathrm{ ratio}$ and Eddington ratio. This lack of dependence of $d_\mathrm{ratio}$ on Eddington ratio can be understood as a result of cancellation between the two previous anticorrelations. 

A promising explanation for the longer-than-expected inter-band lags is that the diffuse continuum emission from the BLR makes a substantial contribution to optical continuum variability \citep{Korista01,Lawther18,Chelouche19,Netzer21,Guo22}. This idea is strongly supported by the $u/U$-band excess observed in the lag spectrum of several objects, although it is not shown in every nearby AGN \citep[e.g., Mrk~817,][]{Kara21}. The optical bands are strongly contaminated by the free-bound Balmer and Paschen continua, especially in spectral regions at $\sim$3700 \AA\ and 8200~\AA. Based on multicomponent decompositions of AGN spectra around the  Balmer and Paschen edges of several nearby AGN, the fraction of the diffuse BLR emission flux relative to the total continuum is found to be around 10\% to 50\% \citep{Vincentelli21,Guo22}, in general agreement with the predictions of BLR photoionization models. \citep[e.g., NGC 5548, Fairall 9, Mrk 110,][]{Lawther18,HernandezSantisteban20,Vincentelli21}. Although the flux of the diffuse BLR emission may not be dominant ($<$50\%) in the continuum emission, its lag contribution could still be primary as the lag of the BLR emission is expected to be much longer than that of disk emission. Further work on modeling the combined time-variable contributions of diffuse BLR emission and disk emission \citep[see, e.g.,][]{Jaiswal22} is still needed in order to fully determine the BLR contribution to broad-band continuum emission. Even in scenarios where lamp-post reprocessing is not the primary cause of disk variability (such as  models in which thermal fluctuations drive disk variability, as described in \S \ref{sec:intro}), the diffuse BLR emission would still be expected to make a substantial contribution to the overall optical continuum and lags. 

If diffuse BLR emission indeed makes a major contribution to the continuum lags, the observed anti-correlation of disk-size ratio and luminosity can be naturally explained by a diffuse BLR Baldwin effect, i.e., an anti-correlation of diffuse BLR emission equivalent width and continuum luminosity \citep[also see][]{Li21}. As both the diffuse BLR emission and quasar emission lines (such as \ion{He}{2} $\lambda$1640) are dominated by recombination emission from the BLR, a diminished contribution from diffuse emission is expected with increasing central luminosity, thus reducing the diffuse BLR lag with respect to disk lag and alleviating the disk size discrepancy at higher luminosities. The intrinsic origin of the Baldwin effect is still under debate. Photoionization modeling attributes it to a characteristic relationship between the continuum spectral energy distribution, the gas metallicity, and the quasar luminosity \citep{Korista99}, while other work has suggested that the Eddington ratio is the intrinsic driver \citep{Dong09}. The anti-correlation of $d_\mathrm{ratio}$ with $M_\mathrm{BH}$ can be introduced by the $L \propto M_\mathrm{BH}$ relation of our sample (see \S \ref{sec:r-L}), although the Spearman correlation coefficients indicate that its correlation is stronger than that of $d_\mathrm{ ratio}-L$.

Interestingly, \citet{Vincentelli22}, together with \citet{Vincentelli21}, observed an opposite trend in two intensive reverberation monitoring campaigns on Mrk 110. They divided two campaigns into three intervals based on X-ray luminosity, and found that the $U$-band lag excess is not observed in the lowest X-ray state but is seen in two subsequent intervals with higher X-ray luminosities, which indicates an opposite luminosity-dependent diffuse BLR lag in Mrk~110. They argued that the increase of $U$-band lag with luminosity could be naturally explained by an increasing radius of the diffuse BLR emission induced by a rising ionizing continuum luminosity.

\begin{figure}
\hspace{-1cm}
 \centering
   \includegraphics[width=0.5\textwidth]{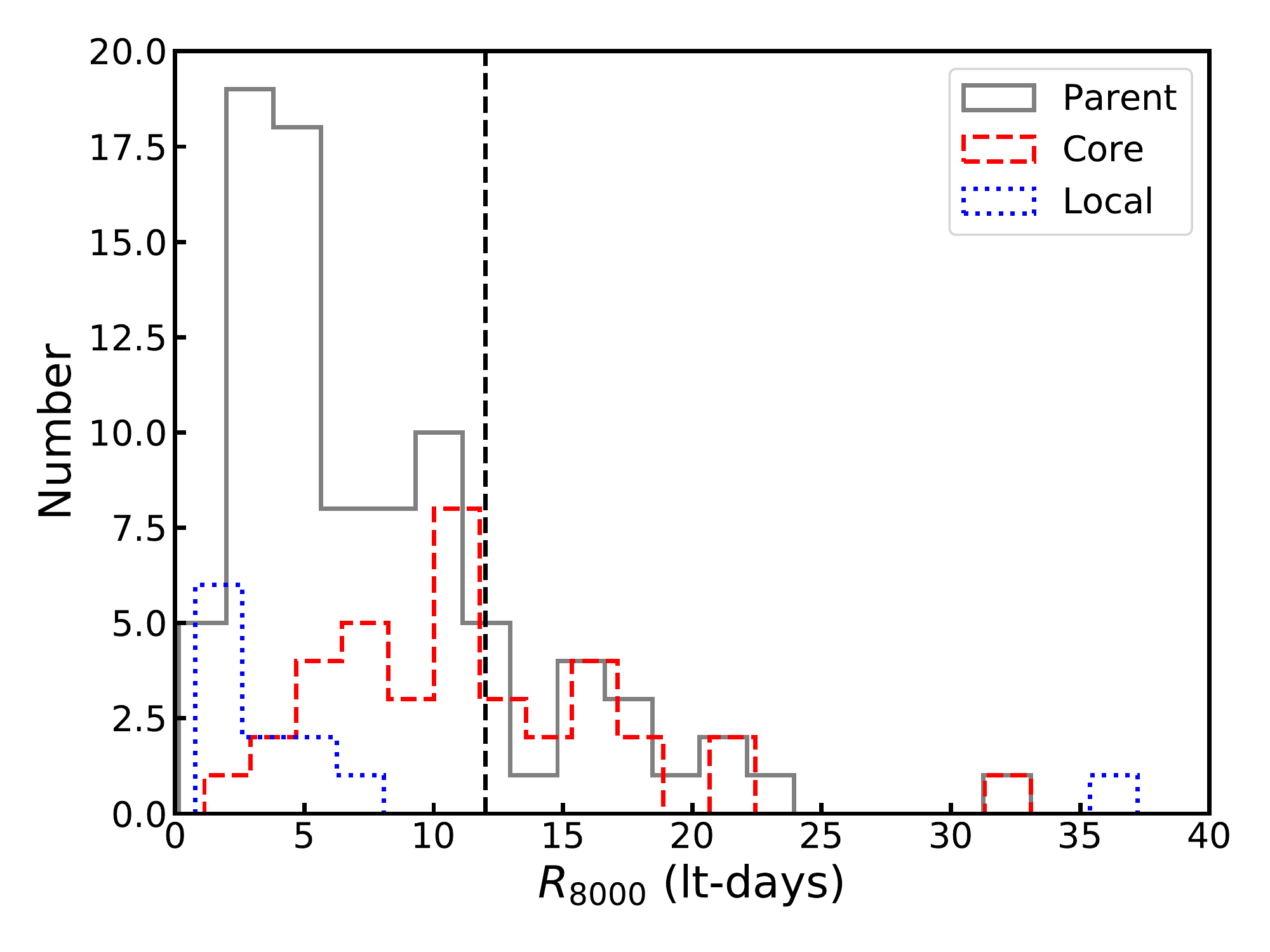}
    \caption{Distributions of CE size at 8000 \AA\ for different samples. The self-gravity limit of $R_\mathrm{ sg}$ = 12 days is indicated by the black dashed line. Different samples are labeled with different lines.}
    \label{fig:gravity}
\end{figure}

\begin{figure*}
 \centering
 \includegraphics[width=1\textwidth]{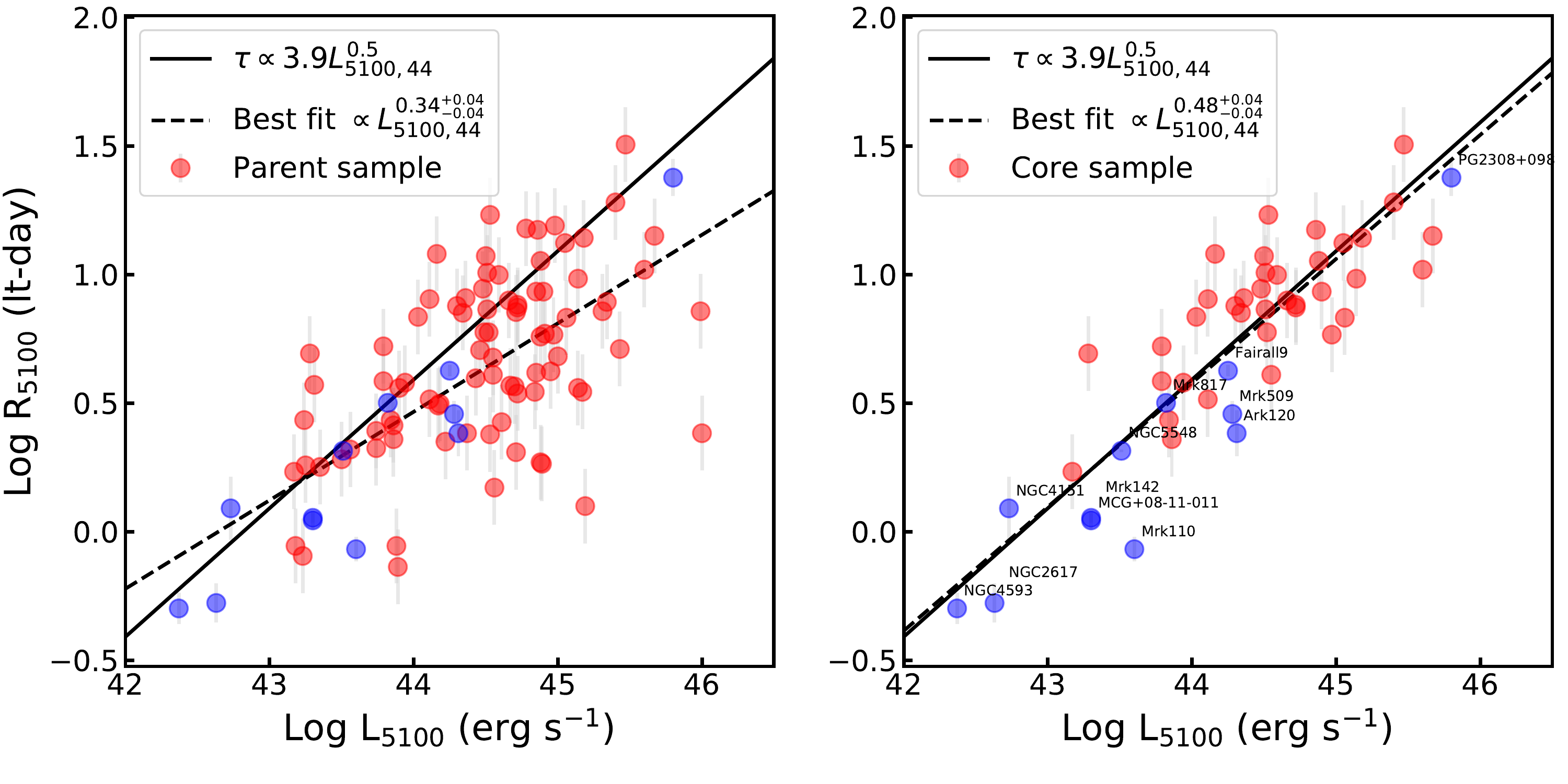}
    \caption{CE size as a function of continuum luminosity at 5100 \AA. Lag measurements and 1$\sigma$ errors from other low-redshift continuum RM campaigns (see Table \ref{tab:localAGN}) are also shown. The best linear MCMC fit is based on the combined sample assuming a typical luminosity uncertainty of 0.05 dex. The $\tau-L$ relation predicted by the RPC model from \citet{Netzer21} is also displayed without any normalization. }
    \label{fig:lag-L}
\end{figure*}

\begin{figure*}
 \centering
   \includegraphics[width=1\textwidth]{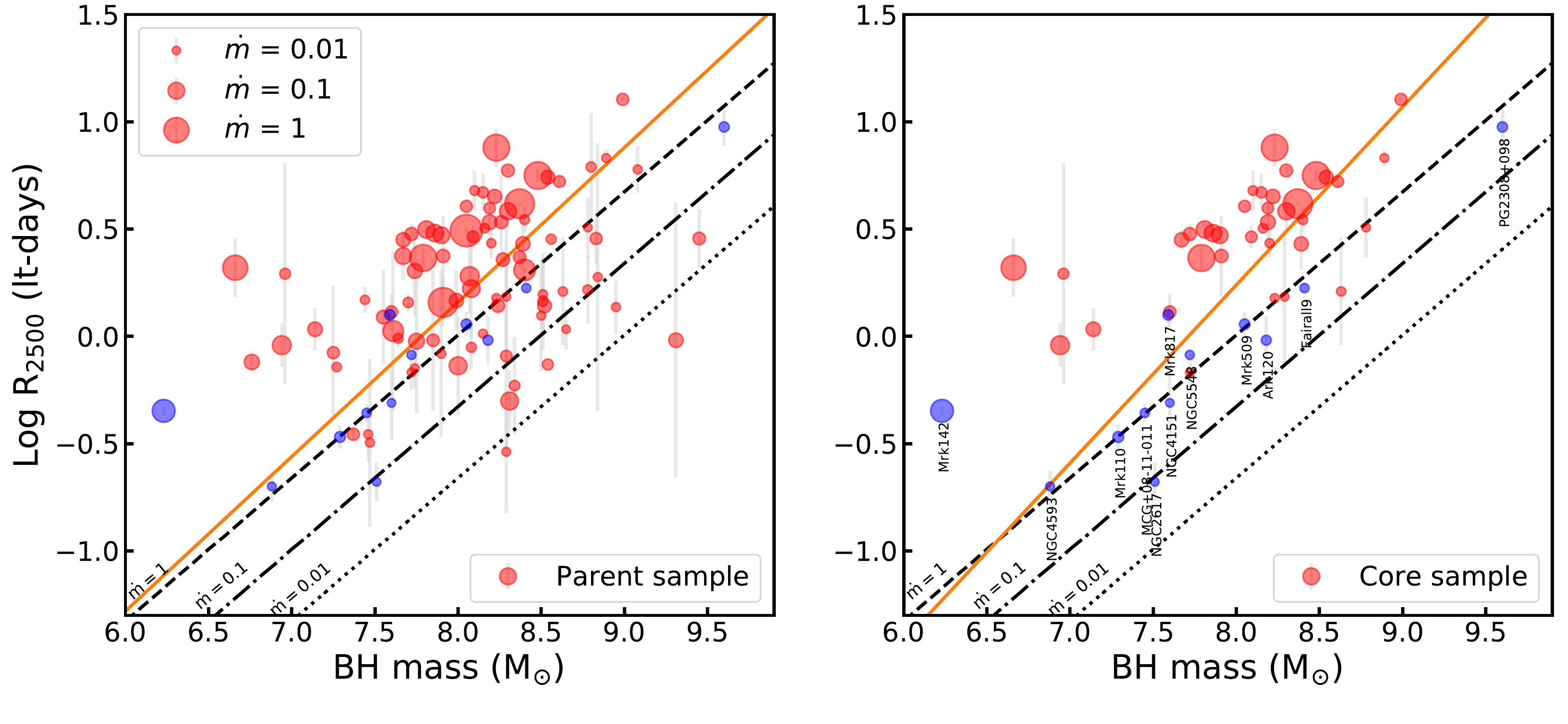}
    \caption{CE size at 2500 \AA\ as a function of BH mass. Red and blue dots with 1$\sigma$ errors denote the ZTF sample and local AGN sample, respectively. Dots with different sizes present the corresponding Eddington ratios and three examples of $\dot{m}=$ 0.01, 0.1, and 1 are listed. Different lines (with slope of 2/3 assuming $L \propto M_{\mathrm{BH}} $ following \citet{Yu20}, slightly different from the real situation) indicates the predictions from the SSD model with Eddington ratios $\dot{m}=$ 0.01, 0.1, and 1. The orange lines are the best linear fits for combined samples with slopes of $0.72$ (left) and $0.83$ (right), assuming the typical error of BH mass is 0.4 dex.}
    \label{fig:lag-BH}
\end{figure*}

\subsection{Accretion Disk Self-Gravity Limit}\label{sec:gravity}

Models of accretion disk structure predict that vertical self-gravity becomes important at large radii, and the disk is expected to be fragment beyond a certain radius (i.e., a self-gravity limit, $R_\mathrm{ sg}$). This provides a natural outer boundary for the disk, $R_\mathrm{ out}$. A remarkable feature in standard thin disk is that the self-gravity radius is roughly constant, almost independent of other parameters according to  \citet{Collin-Souffrin90} and \citet{Lobban22}:
\begin{equation}
    R_\mathrm{sg} \simeq 3 \times 10^{16}\ \textrm{cm} \simeq 12\ \textrm{lt-days}.
\end{equation}
This predicts the disk size measured from reverberation should be bounded by an upper limit of about 12 lt-days, if the continuum reverberation lags are dominated by the disk contribution and not by the BLR contribution originating on larger scales. 

To compare with this theoretical prediction, we extrapolate the CE size from 2500 \AA\ to 8000 \AA\ assuming $R_\mathrm{ \lambda} \propto \lambda^{4/3}$, which corresponds to T$\sim$3,600 K according to the Wien law. We emphasize this is a lower limit of $R_{\mathrm{out}}$ and the real outer boundary can extend further. Figure \ref{fig:gravity} demonstrates that a large fraction of AGN in either the parent sample ($\sim$30\%) or core sample ($\sim$60\%) have CE radii at 8000 \AA\ that exceed the self-gravity limit of 12 lt-days. This strongly supports the idea that the diffuse BLR component significantly contributes to the continuum lags. The fact that the AGN from the local sample have CE sizes below this limit can be explained by their low luminosities. Alternatively, a Corona-Heated Accretion-disk Reprocessing (CHAR) model \citep{Sun20a}, which assumes the variability is caused by temperature fluctuations in the accretion disk and the corona is coupled to the magnetic field, can also provide an explanation for the inferred CE sizes exceeding the disk self-gravity limit, since the time lags in CHAR model are caused by the radius-dependent thermal timescales which can be much longer than the light-travel time delays at the self-gravity radii.

\subsection{Lag - Luminosity Relation}\label{sec:r-L}
According to the locally optimally emitting cloud (LOC) photoionization model, which posits clouds of different gas densities distributed over a range of distances from the central continuum source \citep{Baldwin95}, the line emission we observe originates from the combination of all clouds. For a given line the emission is dominated by those clouds with the highest efficiency of reprocessing the incident ionizing continuum into that line, i.e., these clouds with the optimal distance from the central source and gas density. For a BLR cloud, the ionization parameter $U_{\mathrm{H}}$ is given by
\begin{equation}
    U_{\mathrm{ H}} = \frac{Q(H)}{4\pi R^{2}n_{\mathrm{ H}}c},
\end{equation}
where $Q(H)$ is the rate of production of hydrogen-ionizing photons by the AGN, and $n(H)$ is the hydrogen density. As $L\propto Q(H)$, this yields $R \propto L^{0.5}$ under the assumptions that the ionization parameters and particle densities are about the same for all AGN. The well-known \hb\ $R-L$ relation \citep{Bentz13} is observed to follow this scaling. Likewise, if the continuum lag is dominated by the BLR diffuse emission, a similar relationship is expected. 

Otherwise, if the inter-band lags are dominated by the disk components, the lags are expected to be proportional to $(M_{\mathrm{ BH}}L)^{1/3} $ (see Equation \ref{eq:R}). This can be reduced to a relation between $\tau$ and $L$ by using the mass-luminosity relation of the sample to eliminate $M_\mathrm{BH}$ from the scaling. We performed a linear fit of the $L-M_\mathrm{BH}$ relation and obtained a relationship of $L \propto M_{\mathrm{BH}}^{1.11}$ with a scatter of 0.1 dex for both the P$+$L and C$+$L sample. This indicates that for our sample $\tau$ should scale with $L^{0.63}$ if the disk model of Equation \ref{eq:R} applies  \citep[also see the similar discussion in][]{Montano22}.

Figure \ref{fig:lag-L} displays the $R_{\mathrm{CE}}-L$ relation at 5100~\AA\ for both the parent and core sample. The local AGN with more reliable CE-size measurements due to the intensive sampling and long-baseline light curves are also added in the plots (see Table \ref{tab:localAGN}). All CE sizes at 5100 \AA\ are extrapolated from $R_{2500}$ with $R_{\mathrm{ \lambda}} \propto \lambda^{4/3}$. Clearly, there is a positive correlation in the C$+$L sample whereas the correlation is more ambiguous in the P$+$L sample. This is because criterion 1 in parent sample selection includes some objects with less reliable lag detections (i.e., $R_{\mathrm{5100}} < 3$ lt-days) and the potential Baldwin effect of the diffuse BLR emission, thus increasing the scatter and biasing the relationship. Therefore, we consider the results from the C$+$L sample to be more reliable than those from the P$+$L sample.

In order to parameterize the $R_{\mathrm{CE}}-L$ relation, we apply a linear MCMC fitting for the combined samples using the python package {\tt linmix} \citep{Kelly07}, by fitting the function
\begin{equation}
    \log({R}_{\mathrm{5100}}/\textrm{1\ lt-day}) = \alpha\ \log({L}_{\mathrm 5100}/10^{44}\ \textrm{erg\ s}^{-1} ) + K.
\end{equation}
The best fit to the data gives $\alpha = 0.48^{+0.04}_{-0.04}$, $K = 0.56^{+0.03}_{-0.03}$ ($\alpha =0.31^{+0.04}_{-0.04}$, $K = 0.48^{+0.03}_{-0.03}$) for the core (parent) sample. The scatter of the $R_\mathrm{CE}-L$ relation from the linear fit is $\sim$0.2 (0.5) dex for core (parent) sample. We find that the fiducial $R_\mathrm{CE}-L$ relation in C$+$L sample is indeed similar to the well-known \hb\ $R-L$ relation, and differs from the slope of 0.63 predicted by the SSD model as described above. This result is also consistent with the slope of $\sim$0.4$-$0.5 discovered by \citet{Sergeev05} and also confirmed by \citet{Montano22}, which extends to a much lower luminosity regime by successfully detecting the continuum lag of the low-mass AGN in NGC 4395.

The right panel of Figure \ref{fig:lag-L} seemingly shows that the sample distribution is in good agreement with $R_\mathrm{CE} \propto L^{1/2}$, and the diffuse BLR contribution to continuum may be similar over 4 orders of magnitude in continuum luminosity at 5100~\AA. However, Figure \ref{fig:disk_ratio} implies a potential evolution of $d_{\mathrm {ratio}}$ as a function of monochromatic luminosity (a potential Baldwin effect of diffuse BLR emission), which means we should expect to see luminosity-dependent trends with respect to the non-evolutionary $R \propto L^{1/2}$ for sub-Eddington accreting AGN. Indeed, we found that the best-fit slope is slightly flatter than 0.5, which reflects an expected mild anti-correlation of $d_{\mathrm {ratio}}$-L in C$+$L sample. This evolution effect is more obvious in P$+$L sample (see the top left panel of Figure \ref{fig:disk_ratio}), thus yielding a much flatter slope in $R_\mathrm{CE}-L$ relation. This indicates the Baldwin effect of diffuse BLR emission and the $R_\mathrm{CE}-L$ relation are well connected. Nevertheless, the strength of the anti-correlation between $d_{\mathrm {ratio}}$ and $L$ is still indeterminate limited by the sample size and ambiguous lag detections, and it is more challenging to quantify its effect on the $R_\mathrm{CE}-L$ relation.  

To further examine the physical implications of these measurements, we introduce a newly established $\tau-L$ relation based on a radiation pressure confined BLR cloud (RPC) model from \citet{Netzer21}. The RPC model also works in the framework of the lamp-post reprocessing scenario. Its diffuse BLR emission modeling follows \citet{Lawther18} and \citet{Korista19}. According to their model, a first-order approximation of the total observed continuum lag at wavelength $\lambda$ is:
\begin{equation}\label{eq:lag}
    \tau_{\mathrm{ \lambda, total}} = \tau_\mathrm{\lambda, disk}\frac{L_\mathrm{ incident}}{L_\mathrm{ total}} + \tau_\mathrm{ \lambda, diff}\frac{L_\mathrm{ diff}}{L_\mathrm{ total}} \ \rm{days},
\end{equation}
where $\tau_\mathrm{\lambda, disk}$ and $\tau_{\mathrm{\lambda, diff}}$ refer to the lags of variation in disk and diffuse BLR emission relative to the central variability. If the disk lag ($\tau_\mathrm{ \lambda, disk}$) is negligible and we further assume the diffuse BLR continuum lag ($\tau_\mathrm{ \lambda, diff}$) to be around half of the broad \hb\ lag \citep[e.g.,][]{Korista19,Netzer21},
\begin{equation}
    \tau_\mathrm{ \lambda, diff} \simeq 0.5\tau_\mathrm{ H\beta} = 17L^{0.5}_{5100,44} \ \rm{days},
\end{equation}
then the continuum lag at 5100 \AA\ can be written as: 
\begin{equation}
    \tau_{5100} \approx 17L_{5100,44}^{0.5}\frac{1.5c_{f}}{1+1.5c_{f}} \ \rm{days},
\end{equation}
where $c_{f}$ is the BLR covering factor, and $L_\mathrm{ 5100,44}$ represents $L_\mathrm{ 5100}$ in units of $10^{44}\ \rm erg\ s^{-1}$. \citet{Netzer21} found that the diffuse continuum from a BLR with a covering factor of $c_{f} = 0.2$ (i.e., $L_\mathrm{ diff,5100}/L_\mathrm{ total} \simeq 0.23$ at 5100 \AA\ and $\tau \propto 3.9L_{5100,44}^{0.5}$ days) can explain the entire observed lag spectra in terms of both shape and magnitude, without necessarily requiring any lag contribution from the disk. Furthermore, the $\tau-L$ relation in local AGN is in good agreement with the prediction ($\tau \propto 3.9L_{5100,44}^{0.5}$ days, see their Figure 4). However, the sample used by \citet{Netzer21} is limited in size ($<$ 10 AGN) and the luminosity range of the sample is limited to objects with $\lesssim 10^{44}$ erg~s$^{-1}$.

In this work, we further examine this relationship with a much larger sample size and luminosity range ($10^{42}$ to $10^{46}$ erg~s$^{-1}$). Remarkably, we find that the $\tau-L$ relation ($\tau = R_{\mathrm{CE}}/c$) predicted by the RPC model closely matches the C$+$L sample data, as has also been found by extending the sample to the much lower luminosity of NGC 4395 \citep{Montano22}. This provides further corroborating evidence that the diffuse BLR emission has a strong influence on AGN continuum lags. However, we cannot rule out the standard thin disk model if  considering the monochromatic luminosity in a self-similar scenario as it also produces a slope of 0.5 in the $\tau-L$ relation (see details in Appendix \ref{app:self-similar}). But we suggest that the BLR origin of the $\tau-L$ relation is more compatible with other observational facts, for instance, the \emph{U}-band excess caused by Balmer continuum emission from BLR clouds seen in NGC~4593 \citep{Cackett18} and other nearby AGN having high-cadence continuum reverberation mapping data \citep[e.g.,][]{Fausnaugh16,Edelson19}.   Furthermore, it implies that the average diffuse BLR continuum fraction at 5100 \AA\ in our sample is around 23\% with respect to the total luminosity if the RPC model is approximately correct. At the wavelengths of the Balmer and Paschen jumps, this fraction can be as high as 30\% to 40\%, predicted by \citet{Korista19} (see their Figure 9), consistent with the results from \S \ref{sec:disksize} and \ref{sec:gravity}. Finally, this model indicates the $\tau-L$ relation for the observed continuum lag (a superposition of disk and diffuse BLR lags) is similar to  the \hb\ $R-L$ relation but down-scaled by a factor of 8.7 (17$\times$2/3.9) in normalization. In the future, more complex BLR modeling \citep[e.g.,][]{Lawther18} combined with a spectral decomposition of high S/N spectra covering a broad wavelength range \citep[e.g.,][]{Guo22} is still needed to disentangle the contributions from disk and diffuse BLR emission and to obtain estimates of the intrinsic disk lags.

\subsection{CE size - BH mass Dependence}\label{sec:r-m}
Analogous to the $R-L$ dependence, the SSD model also predicts a positive correlation between the accretion disk size and the BH mass: $\tau \propto (M_\mathrm{ BH}L)^{1/3} \propto M_\mathrm{ BH}^{0.70}$ based on the scaling of $L \propto M_{\mathrm{BH}}^{1.11}$ in both the P$+$L and C$+$L samples.

Figure \ref{fig:lag-BH} displays the CE size at 2500~\AA\ as a function of BH mass. Significant positive correlations are obtained in both panels, with Spearman correlation coefficient $r = 0.41$ and $p =1\times10^{-5}$ ($r = 0.63$, $p =8\times10^{-7}$) for the P$+$L (C$+$L) sample. The linear MCMC fit shows that the slopes are $0.72^{+0.12}_{-0.13}$ and $0.83^{+0.13}_{-0.16}$ for the P$+$L and C$+$L samples, considering a 0.4 dex uncertainty of BH mass in the linear regression. We conclude that the positive correlation is robust, whereas the slope is not well constrained, yet not significantly different from the SSD prediction (0.70). The scatter of this relation is relatively large ($\sim 1$ dex), which is mainly due to the diverse values of $\dot{m}$ for the sample, probably indicating different accretion modes and disk geometries. According to the SSD expectation, AGN with higher accretion rates will yield larger CE size given the same BH mass (see the theoretical CE sizes $R_{\mathrm{2500}}$ as a function of BH mass with different Eddington ratios, three black lines in Figure \ref{fig:lag-BH}.) The commonly observed larger CE sizes in our sample with respect to SSD prediction at corresponding accretion rates reflects the unexpected larger disk size in \S \ref{sec:disksize}.

Compared with previous detections of weak trends between CE size and BH mass \citep[e.g.,][]{Yu20,Jha21,Guo22}, the positive correlation in this work is relatively more robust for three reasons: first, our lag selection criteria are stricter and our sample size is relatively larger; secondly, local AGN with low BH masses are considered in our analysis, which extends the BH mass range and better constraints the relationships; finally, the high-quality light curves (high cadence and high S/N) together with the intrinsic properties of the AGN (short-term variability timescales for low-redshift objects) further reinforce the short-lag detection ability.

\section{Conclusions}\label{sec:con}
We compiled a sample of low-redshift ($z < $0.8) AGN with robust lag measurements by cross-matching the MQC type-1 AGN with ZTF light curves to constrain the CE size. Two inter-band lags, $\tau_\mathrm{ gr}$ and $\tau_\mathrm{ gi}$, were measured using the standard ICCF method and the Bayesian code {\tt JAVELIN}. We made a strict sample selection,  requiring a relatively high cross-correlation $r_\mathrm{ max}> 0.8$,  consistency in lag between two methods, and small lag uncertainties. The selection procedure yields a parent sample of 94 AGN and a core sample of 38 AGN with different lag quality. These measurements were combined with published results for local AGN from targeted continuum RM campaigns. Our main findings are as follows:
\begin{enumerate}
    \item Our results suggests that the increase in continuum lags from $g-r$ to $g-i$ is generally consistent with the SSD prediction with a slope of 4/3. The CE size is larger than disk size predicted in the SSD model by a factor of $\sim$3 with $X=2.49$, similar to the results from disk RM and microlensing.  
    
    \item Self-gravity theory predicts that the accretion disk size has an upper limit of around 12 lt-days (the self-gravity limit), almost independent of any parameters \citep{Lobban22}. However, our measurements show that a significant fraction (30\% to 60\%) of the AGN in our sample have CE sizes at 8000 \AA\ exceeding this limit, which can be interpreted as evidence for a significant contribution from the diffuse BLR emission to the continuum lags.  
    
    \item The ratio between CE size and disk size predicted in the SSD model anti-correlates both with the continuum luminosity and the BH mass. \citet{Li21} claimed that this may be caused by a Baldwin effect of the diffuse BLR emission, analogous to normal quasar emission lines (e.g., \ion{He}{2}): with increasing continuum luminosity, the diffuse BLR contribution diminishes, thus reducing the diffuse BLR lag relative to the accretion disk lag and alleviating the disk size discrepancy.  
    
    \item The CE size at 5100 \AA\ scales with the continuum luminosity as $R_{\mathrm{CE}} \propto L^{0.48\pm0.04}$ with a scatter of 0.2 dex, very similar to the $R-L$ relation of the broad \hb\ line. This again provides important evidence that the observed continuum lags contain a significant contribution from the diffuse BLR continuum (with an expected $R\propto L^{0.5}$). Moreover, this $R_{\mathrm{CE}}-L$ dependence also closely matches a radiation pressure confined BLR cloud model, which assumes that the disk lag is negligible and the diffuse continuum emission originates from a BLR with a covering factor of 0.2 \citep{Netzer21}. If this is the case, it indicates the average fraction of the diffuse BLR emission relative to the total flux is around 23\% at 5100 \AA\ in our sample.
    
    \item A robust positive correlation between the CE size and the BH mass is found. However, the slope of this relationship is not well constrained due to the diverse accretion rates.
\end{enumerate}

If there is indeed a significant contribution of diffuse BLR emission to continuum lags that decreases as a function of luminosity, this would also produce a luminosity-dependent bias in broad emission-line reverberation lags. H$\beta$ lags are usually measured relative to the optical continuum in the $g$-band or $V$-band spectral region, as a proxy for the unobservable ionizing continuum, and a contribution of diffuse BLR emission to the optical continuum would tend to decrease the measured emission-line lags. Any luminosity dependence to the DC emission fraction would then affect the slope of the observed H$\beta$ radius-luminosity relationship \citep{Bentz13}. The impact on virial BH  masses determined from RM might be modest, since the virial normalization factor (the $f$-factor) is externally calibrated based on the $M_\mathrm{BH}-\sigma$ relation \citep{Onken2004}. However, dynamical modeling methods for determination of BH masses from RM data \citep[e.g.,][]{Pancoast2011} generally assume that the observed optical continuum emission originates from a region of negligible size compared with the BLR radius, and these methods may need revision to account for a substantial (and luminosity-dependent) contribution of BLR continuum emission.

With further release of ZTF $g$- and $r$-band light curves, more lags will be reliably detected in AGN and it will also allow us to explore the lag variations on timescale of a few years. However, despite the ZTF light curves being generally better than other surveys, the ability to detect lags in the survey data is still limited, for example, by the cadence and S/N for very short lags, and by weak variability in many light curves. In the future, the Vera Rubin Observatory will provide higher cadence (e.g., in the Deep Drilling Fields) and more accurate light curves with wider sky coverage and additional filter bands \citep[e.g.,][]{Brandt18,Ivezic19,Kovacevic22}. It will significantly push forward our outstanding of the accretion disk physics and variability mechanism in AGN, and distinguish the X-ray reprocessing scenario from other pictures \citep[e.g., the CHAR model,][]{Sun20a}.

\begin{acknowledgements}
We thank the anonymous referee for providing helpful suggestions. We thank Mouyuan Sun for helpful discussion about the CHAR model. We thank Kirk T. Korista and Michael R. Goad for providing the diffuse BLR model and helpful discussion. Research at UC Irvine has been supported by NSF grant AST-1907290.  Shu Wang is supported by the National Research Foundation of Korea (NRF) grant funded by the Korean government (MEST) (No. 2019R1A6A1A10073437)

Based on observations obtained with the Samuel Oschin 48 inch Telescope at the Palomar Observatory as part of the Zwicky Transient Facility project. ZTF is supported by the National Science Foundation under grant No. AST-1440341 and a collaboration including Caltech, IPAC, the Weizmann Institute for Science, the Oskar Klein Center at Stockholm University, the University of Maryland, the University of Washington, Deutsches Elektronen-Synchrotron and Humboldt University, Los Alamos National Laboratories, the TANGO Consortium of Taiwan, the University of Wisconsin at Milwaukee, and Lawrence Berkeley National Laboratories. Operations are conducted by COO, IPAC, and UW.

Funding for the Sloan Digital Sky Survey IV has been provided by the Alfred P. Sloan Foundation, the U.S. Department of Energy Office of Science, and the Participating Institutions. 

SDSS-IV acknowledges support and resources from the Center for High Performance Computing  at the University of Utah. The SDSS website is www.sdss.org.

SDSS-IV is managed by the Astrophysical Research Consortium for the Participating Institutions of the SDSS Collaboration including the Brazilian Participation Group, the Carnegie Institution for Science, Carnegie Mellon University, Center for Astrophysics | Harvard \&  Smithsonian, the Chilean Participation Group, the French Participation Group, Instituto de Astrof\'isica de Canarias, The Johns Hopkins University, Kavli Institute for the Physics and Mathematics of the Universe (IPMU) / University of Tokyo, the Korean Participation Group, Lawrence Berkeley National Laboratory, Leibniz Institut f\"ur Astrophysik Potsdam (AIP),  Max-Planck-Institut f\"ur Astronomie (MPIA Heidelberg), Max-Planck-Institut f\"ur Astrophysik (MPA Garching), Max-Planck-Institut f\"ur Extraterrestrische Physik (MPE), National Astronomical Observatories of China, New Mexico State University, New York University, University of Notre Dame, Observat\'ario Nacional / MCTI, The Ohio State University, Pennsylvania State University, Shanghai Astronomical Observatory, United Kingdom Participation Group, 
Universidad Nacional Aut\'onoma de M\'exico, University of Arizona, University of Colorado Boulder, University of Oxford, University of  Portsmouth, University of Utah, University of Virginia, University of Washington, University of Wisconsin, Vanderbilt University, and Yale University.
\end{acknowledgements} 

\clearpage
\facility{ZTF\citep{https://doi.org/10.26131/irsa539}, SDSS}
\software{AstroPy \citep{Astropy2018}, PyI$^2$CCF \citep{Guo21b}, Linmix \citep{Brandon07}}

\appendix
\renewcommand\thefigure{\thesection.\arabic{figure}} 
\setcounter{figure}{0}

\section{An AGN with a temporally abnormal time delay}\label{app:abnormal}
\begin{figure*}
\hspace{-1cm}
 \centering
 \includegraphics[width=16cm]{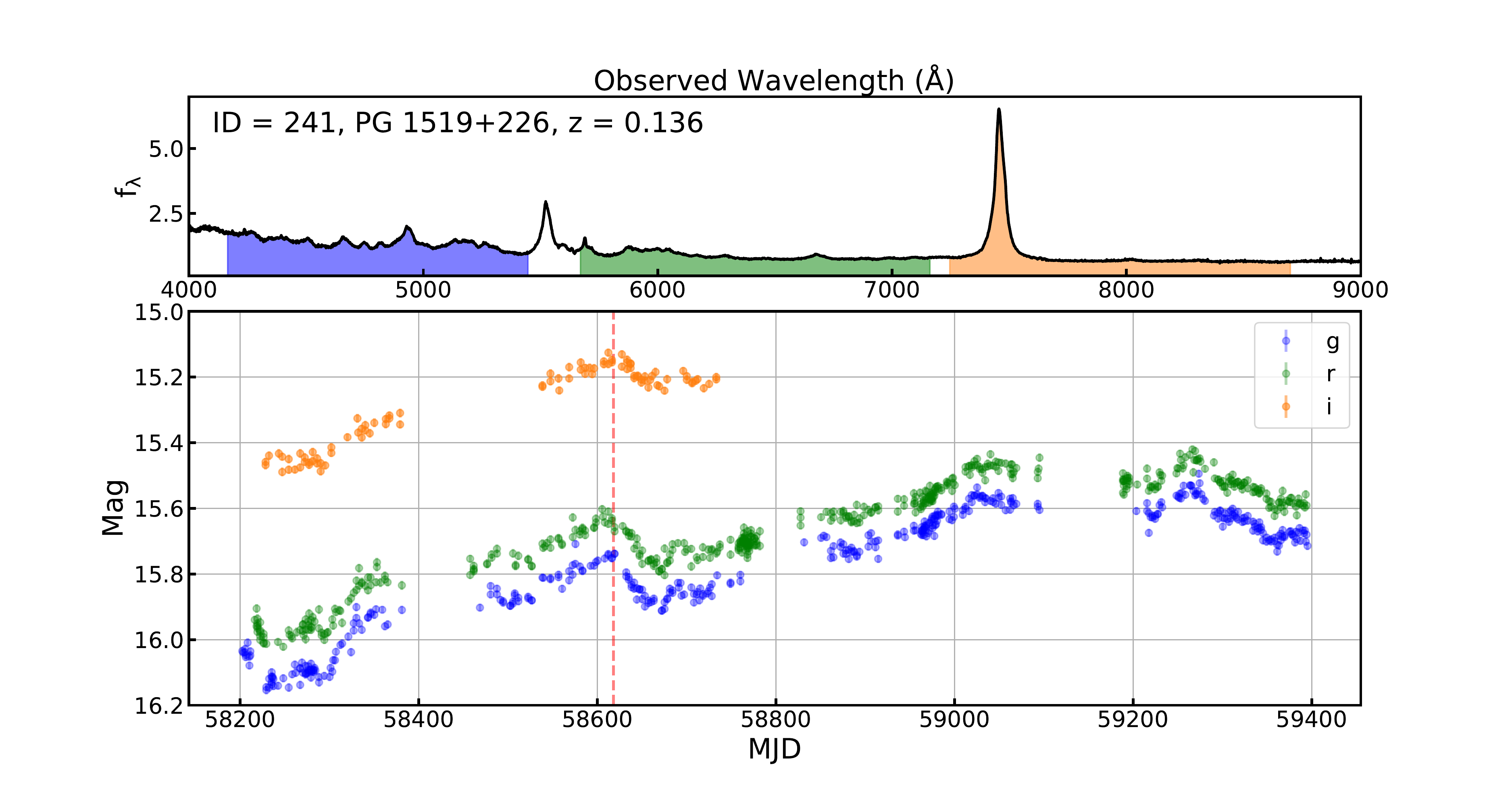}
    \caption{Upper panel: SDSS spectrum overlapped with the ZTF $gri$ filters. Lower panel: ZTF light curves showing a temporally negative $\tau_\mathrm{ gr}$ ($r$-band leads $g$-band) around MJD = 58600. The red dashed line indicates the $g$-band peak, obviously lagging to that of $r$-band. Also see the overall lag measurements in the first panel of Figure \ref{fig:example}.}
    \label{fig:r-g.pdf}
\end{figure*}

Here we describe an AGN (ID = 241, PG 1519$+$226 at $z=0.136$) exhibiting unexpected behavior in which the $r$-band light curve temporally leads the $g$ band (as well as the $i$ band) by a few days ($<$ 5 days). This is most evident by inspection of a peak in the light curve around MJD $=$ 58600, as illustrated in Figure \ref{fig:r-g.pdf} (also see its overall lag measurements in the first panel of Figure \ref{fig:example}), where the $r$-band peak occurs noticeably earlier than the $g$-band peak. We also performed a visual inspection for other objects in the initial sample of 455 AGN and did not find any other objects showing evidence of similar behavior, indicating a detection rate of $\lesssim 0.2$\%. An inverted relationship between lag and wavelength is not compatible with the standard disk reprocessing model, but may be explained by other scenarios for the origin of continuum variability. As indicated by the inhomogeneous accretion disk model \citep[e.g.,][]{Dexter11}, if disk temperature fluctuations induce a strong variability amplitude at the optical emitting region, the fluctuations will propagate to both the shorter and longer wavelengths, which could explain the temporal phenomenon here. However, this occasional abnormal behavior is rapidly disappeared in later light curves such that the overall lags measured from the entire light curves are still positive (short-wavelength bands lead the longer-wavelength bands), as shown in the first panel of Figure \ref{fig:example}. If this variability mechanism is important, we should expect to see more inverted lags in future monitoring. Alternatively, this could also be caused by a significant contamination from broad emission lines in some AGN around a particular redshift since the $g$ and $i$ bands include several emission lines (e.g., \hr\ and Fe emission) whereas the $r$ band does not include any strong emission lines. The negative $r$-band lag could be explained by a strong variability response of the emission lines to continuum variations. Finally, this case should be insensitive to the diffuse BLR emission contribution as the Balmer jump (3646 \AA) barely overlaps with the $g$ band, and the flux of diffuse BLR emission across the $gri$ bands is nearly constant. 

\section{Self-similar scenario in standard thin disk model}\label{app:self-similar}
\begin{figure}
\hspace{-1cm}
 \centering
 \includegraphics[width=16cm]{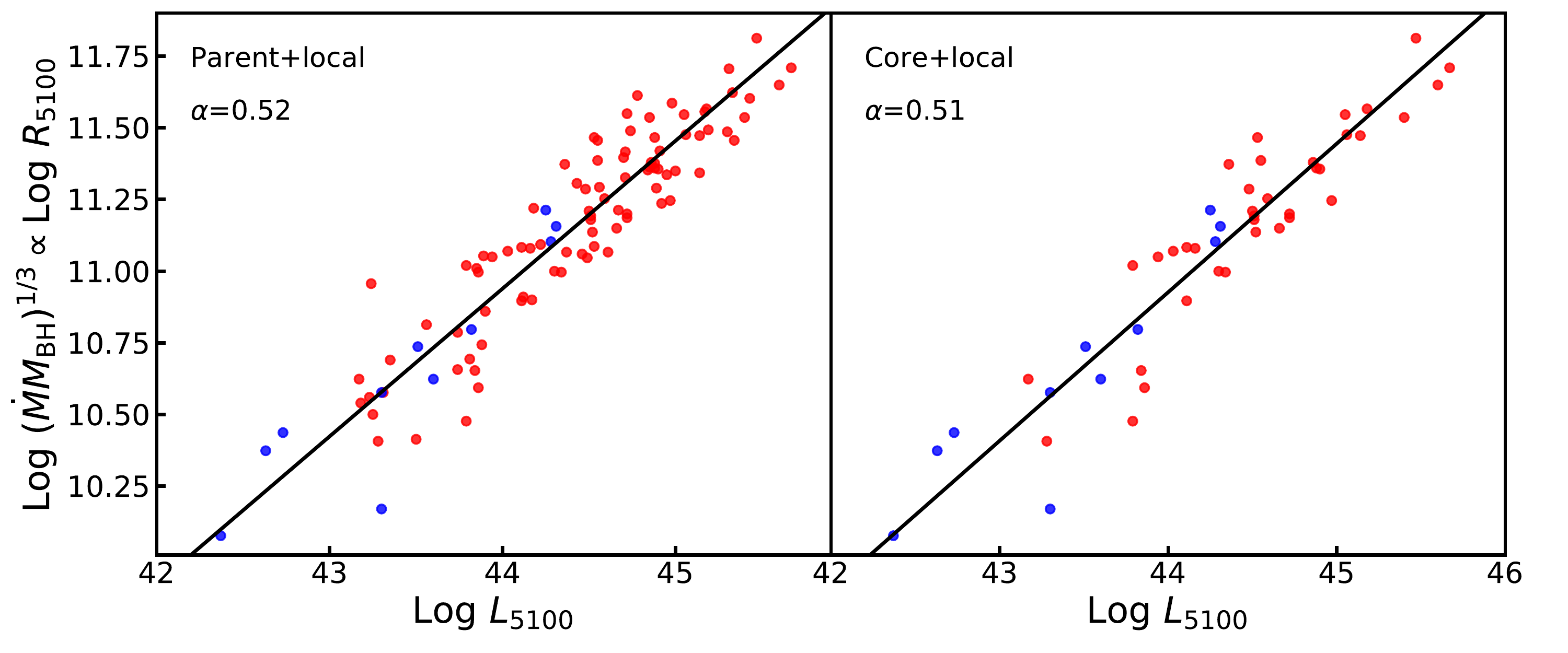}
    \caption{Relationship between monochromatic luminosity at 5100 \AA\ and $\dot{M}^{1/3}M_\mathrm{ BH}^{1/3}$. The blue dots are the local AGN and red dots represent the parent sample and core sample in different panels. The black lines are the best linear fits.}
    \label{fig:self-similar}
\end{figure}

The hypothesis of a BLR origin is not the only possible explanation for a $\tau \propto L^{1/2}$ trend in continuum lags. In the standard thin disk model, when considering a monochromatic luminosity in the self-similar part of the spectral energy distribution, the same slope is predicted. Assuming $T \propto r^{-3/4}$ and assuming that the outer radius is much larger than the inner radius of the accretion disk, the monochromatic luminosity yields $L_{\mathrm\nu} \propto \dot{M}^{2/3}M_\mathrm{ BH}^{2/3}\nu^{1/3}$, where a certain region over the frequency regime can be described by a single power-law between luminosity and frequency ($L_{\mathrm \nu}\propto \nu^{1/3}$), also known as the self-similar regime. Thus, the self-similar scenario predicts $L \propto \dot{M}^{2/3}M_\mathrm{ BH}^{2/3}$. Combining with $R \propto \dot{M}^{1/3}M_\mathrm{ BH}^{1/3}$ derived from Equation \ref{eq:R}, it also predicts a relation of $R \propto L^{1/2}$. Figure \ref{fig:self-similar} exhibits the relation between the observed monochromatic luminosity at 5100 \AA\ and $\dot{M}^{1/3}M_\mathrm{ BH}^{1/3}$ for the P$+$L and P$+$C samples. The linear fits show that the best-fit slopes are indeed close to 0.5, indicating the observed $\tau \propto L^{1/2}$ in the right panel of Figure \ref{fig:lag-L} is also consistent with the self-similar scenario in the standard thin disk model when considering a monochromatic luminosity. However, we argue that the BLR origin is more likely to be the driver of $\tau \propto L^{1/2}$ than the self-similar scenario, as it is more compatible with other observational facts in continuum RM, for example, the $u/U$-band excess observed in several nearby Seyferts having intensive continuum reverberation mapping data \citep{Fausnaugh16,Cackett18,Edelson19}.

\begin{longtable*}{lllll}
\caption{FITS Catalog Format}\label{tab:sample}\\
\hline \hline \\[-2ex]
   \multicolumn{1}{c}{\textbf{Column}} &
   \multicolumn{1}{c}{\textbf{Name}} &
   \multicolumn{1}{c}{\textbf{Format}} &
   \multicolumn{1}{c}{\textbf{Unit}} &
   \multicolumn{1}{c}{\textbf{Description}} \\[0.5ex] \hline
   \\[-1.8ex]
\endfirsthead
1	&ID	     	        &Long           &               &Object ID  \\		
2	&RA	     	        &Double		    &Degree         &Right ascension in decimal degrees (J2000.0) \\	
3	&DEC	            &Double		    &Degree         &Declination in decimal degrees (J2000.0)\\	
4	&Z	     	        &Double		    &               &Redshift cataloged in the Million Quasar Catalog (MQC) \\	
5	&RMAG	     	    &Double		    &Mag            &Red optical magnitude of the object from MQC\\	
6	&NAME	     	    &String		    &               &Designation of the source from MQC\\	
7	&TYPE	     	    &String		    &               &Type classification of the object from MQC\\	
8	&CITE	     	    &String		    &               &Name reference from MQC \\	
9	&ZCITE	     	    &String		    &               &Redshift reference from MQC \\	
10	&CEN\_GR		    &Double	    	&Day            &Observed centroid lag between $g$- and $r$-band from ICCF \\	
11	&CEN\_LOW\_GR	    &Double	        &Day            &Lower measurement error in CEN\_GR  \\	
12	&CEN\_UP\_GR	    &Double	        &Day            &Upper measurement error in CEN\_GR \\	
13	&CEN\_GI		    &Double		    &Day            &Observed centroid lag between $g$- and $i$-band from ICCF \\	
14	&CEN\_LOW\_GI	    &Double	        &Day            &Lower measurement error in CEN\_GI\\	
15	&CEN\_UP\_GI	    &Double	        &Day            &Upper measurement error in CEN\_GI\\	
16	&PEAK\_GR	        &Double		    &Day            &Observed peak lag between $g$- and $r$-band from ICCF \\	
17	&PEAK\_LOW\_GR	    &Double	        &Day            &Lower measurement error in PEAK\_GR \\	
18	&PEAK\_UP\_GR	    &Double	        &Day            &Upper measurement error in PEAK\_GR \\	
19	&PEAK\_GI	        &Double		    &Day            &Observed peak lag between $g$ and $i$-band from ICCF \\	
20	&PEAK\_LOW\_GI	    &Double	        &Day            &Lower measurement error in PEAK\_GI \\	
21	&PEAK\_UP\_GI 	    &Double	        &Day            &Upper measurement error in PEAK\_GI \\	
22	&P\_GR		        &Double		    &               &$P$-value for $g-r$ lag from a cross-correlation reliability test \\	
23	&P\_GI		        &Double		    &               &$P$-value for $g-i$ lag from a cross-correlation reliability test  \\
24	&MED\_GR\_JAV	    &Double	        &Day            &Observed lag between $g$- and $r$- band from {\tt JAVELIN}  \\	
25	&ERR\_LOW\_GR\_JAV	&Double	        &Day            &Lower measurement error in MED\_GR\_JAV \\	
26	&ERR\_UP\_GR\_JAV	&Double	        &Day            &Upper measurement error in MED\_GR\_JAV \\	
27	&MED\_GI\_JAV	    &Double	        &Day            &Observed lag between $g$- and $i$-band from {\tt JAVELIN}  \\	
28	&ERR\_LOW\_GI\_JAV	&Double	        &Day            &Lower measurement error in MED\_GI\_JAV\\	
29	&ERR\_UP\_GI\_JAV	&Double	        &Day            &Upper measurement error in MED\_GI\_JAV \\	
30	&L5100\_L19	        &Double	        &erg s$^{-1}$   &Monochromatic luminosity at 5100 \AA\ from L19\\	
31	&LOGBH\_L19	        &Double	        &M$_\mathrm{ \odot}$&Virial BH mass based on \hb\ from L19 using \citet{Greene05} \\	
32	&L5100\_R20	        &Double	        &erg s$^{-1}$   &Monochromatic luminosity at 5100 \AA\ from R20 \\	
33	&LOGBH\_R20	        &Double	        &M$_\mathrm{ \odot}$&Virial BH mass based on \hb\ from R20 (VP06) \\	
34	&L5100\_S18	        &Double	        &erg s$^{-1}$   &Monochromatic luminosity at 5100 \AA\ from S18 \\	
35	&LOGBH\_S18	        &Double	        &M$_\mathrm{ \odot}$&Virial BH mass based on \hb\ from S18 (VP06) \\	
36	&L5100\_RM	        &Double	        &erg s$^{-1}$   &Monochromatic luminosity at 5100 \AA\ from RM \\	
37	&LOGBH\_RM	        &Double	        &M$_\mathrm{ \odot}$&RM BH mass based on \hb\ from \citet{BentzKatz15} and self-collections\\	
38	&L5100\_LAMOST	    &Double	        &erg s$^{-1}$   &Monochromatic luminosity at 5100 \AA\ from LAMOST \\	
39	&LOGBH\_LAMOST	    &Double	        &M$_\mathrm{ \odot}$&Virial BH mass based on \hb\ from LAMOST DR1 to DR5 (VP06) \\	
40	&L5100\_MEASURED	&Double	        &erg s$^{-1}$   &Monochromatic luminosity at 5100 \AA\ from self-measurements \\	
41	&LOGBH\_MEASURED	&Double	        &M$_\mathrm{ \odot}$&Virial BH mass based on \hb\ (VP06) \\	
42	&L5100		        &Double		    &erg s$^{-1}$   &Adopted fiducial monochromatic luminosity at 5100 \AA \\	
43	&LOGBH		        &Double		    &M$_\mathrm{ \odot}$&Adopted fiducial virial BH mass  \\	
44	&LOG\_EDD	        &Double		    &               &Eddington ratio based on fiducial luminosity and BH mass \\	
45	&R2500\_SSD	        &Double	        &lt-day            &SSD predicted disk size at 2500 \AA\ based on Equation \ref{eq:R} \\	
46	&R2500\_ICCF	    &Double	        &lt-day            &CE size at 2500 \AA\ based on observed inter-band lag modeling  \\	
47	&R2500\_ERR\_ICCF	&Double	        &lt-day            &Measurement error in R2500\_ICCF \\	
48	&R2500\_JAV	        &Double	        &lt-day            &CE size at 2500 \AA\ based on observed inter-band lag modeling \\	
49	&R2500\_ERR\_JAV	&Double	        &lt-day            &Measurement error in R2500\_JAV \\	
50  &FLAG\_SAMPLE       &LONG           &               &Basic sample = 0 \& 1 \& 2; Parent sample = 1 \& 2; Core sample = 2\\
\hline
\end{longtable*}

\bibliography{ref.bib}\label{sec:ref}
\bibliographystyle{aasjournal}

\end{CJK}
\end{document}